\documentclass[useAMS,usenatbib]{mnras}
\usepackage{txfonts}
\usepackage{ae,aecompl}

\usepackage{graphicx}

\title[AIMSS III]{The AIMSS Project -- III. The Stellar Populations of Compact Stellar Systems}

\author[Janz et al.]{Joachim Janz$^{1}$\thanks{E-mail:
jjanz@swin.edu.au}, Mark A. Norris$^{2,3}$, Duncan A. Forbes$^{1}$,  Avon Huxor$^{4}$, \newauthor  Aaron J. Romanowsky$^{5,6}$, 
Matthias J. Frank$^{7}$, Carlos G. Escudero$^{8,9,10}$, \newauthor  Favio R. Faifer$^{8,9,10}$, Juan Carlos Forte$^{10,11}$, Sheila J. Kannappan$^{12}$, 
\newauthor Claudia Maraston$^{13}$, Jean P. Brodie$^{6}$, Jay Strader$^{14}$, Bradley R. Thompson$^{5}$ \\
\\
$^{1}$Centre for Astrophysics \& Supercomputing, Swinburne University, Hawthorn, VIC 3122, Australia\\
$^{2}$Max Planck Institut f\"ur Astronomie, K\"onigstuhl 17, D-69117, Heidelberg, Germany\\
$^{3}$Jeremiah Horrocks Institute, University of Central Lancashire, Preston, PR1 2HE, UK\\
$^{4}$Astronomisches Rechen-Institut, Zentrum f\"ur Astronomie der Universit\"at Heidelberg, M\"onchstra\ss e 12-14, D-69120 Heidelberg,\\
\hskip 0.3cm Germany\\
$^{5}$Department of Physics and Astronomy, San Jos\'e State University, One Washington Square, San Jose, CA 95192, USA\\
$^{6}$University of California Observatories, 1156 High Street, Santa Cruz, CA 95064, USA\\
$^{7}$Landessternwarte, Zentrum f\"ur Astronomie der Universit\"at Heidelberg, K\"onigsstuhl 12, D-69117 Heidelberg, Germany \\
$^{8}$Facultad de Cs. Astron\'omicas y Geof\'isicas, UNLP, Paseo del Bosque S/N, 1900 La Plata, Argentina\\ 
$^{9}$Instituto de Astrof\'isica de La Plata (CCT La Plata - CONICET - UNLP), Paseo del Bosque S/N, B1900FWA La Plata, Argentina \\
$^{10}$Consejo Nacional de Investigaciones Cient\'ificas y T\'ecnicas, Rivadavia 1917, C1033AAJ Ciudad Aut\'onoma de Buenos Aires,\\
\hskip 0.35cm  Argentina \\
$^{11}$Planetario `Galileo Galilei', Secretar\'ia de Cultura, CP1425 Ciudad Aut\'onoma de Buenos Aires, Argentina \\
$^{12}$Department of Physics and Astronomy, UNC-Chapel Hill, CB3255, Phillips Hall, Chapel Hill, NC 27599, USA\\
$^{13}$Institute of Cosmology and Gravitation, Dennis Sciama Building, Burnaby Road, Portsmouth PO1 3FX, UK\\
$^{14}$Department of Physics and Astronomy, Michigan State University, East Lansing, MI 48824, USA
}

\begin{document}

\date{Accepted 2015 November 6.  Received 2015 November 5; in original form 2015 September 30}

\pagerange{\pageref{firstpage}--\pageref{lastpage}} \pubyear{2015}

\label{firstpage}
\maketitle

\begin{abstract}
Over the last decade and a half a growing zoo of compact stellar systems (CSSs) have been found whose physical 
properties (mass, size, velocity dispersion) place them  between those displayed by classical globular 
clusters (GCs) and those of true galaxies. This has led to significant debate about their exact nature.
An important, and until now, underutilized discriminant in this debate is provided by in the stellar population  properties.

Here we present the single stellar population  equivalent ages, metallicities, and [$\alpha$/Fe] 
of 29 CSSs, based on new spectroscopy from 8-10m class telescopes. With the sample compiled from the AIMSS 
project and a search for CSSs in the Sloan Digital Sky Survey (Huxor et al.) we sample CSSs ranging from GCs 
with sizes of merely a few parsec to compact ellipticals (cEs) larger than M32. Together with a literature compilation of 
comparison samples, this provides a panoramic view of the stellar population characteristics of early-type systems.

We find that the CSSs are predominantly more metal rich than typical galaxies at the same 
stellar mass. At high mass, the cEs depart from the mass--metallicity relation of massive 
early-type galaxies, which continuously forms a sequence with dwarf galaxies.  At lower mass, we find a transition 
in the ultracompact dwarf (UCD) metallicity distribution at a few times $10^7$ M$_{\sun}$, which roughly coincides with 
the mass where luminosity function arguments previously suggested the GC population ends. The highest metallicities 
in UCDs and cEs are only paralleled by those of dwarf galaxy nuclei and the central parts of massive early types. 
These findings can be interpreted as an indication that they were more massive at an earlier time and underwent 
tidal interactions to obtain their current mass and compact size. Such an interpretation is supported by CSSs 
with direct evidence for tidal stripping, and by an examination of the CSS escape velocities.\end{abstract}

\section{Introduction}
\defcitealias{AIMSSI}{AIMSS~I}
\defcitealias{AIMSSII}{AIMSS~II}

At the dawn of the millennium, the distinction between star clusters and galaxies was apparently clear.
The parameter space, e.g.~in radius versus mass, between globular clusters (GCs) and galaxies
was essentially empty, allowing simple selections in observed properties to be made in order  to separate star 
clusters and galaxies. The field came to life with the discovery of a population of compact stellar systems 
(CSSs) that started to fill in the parameter space between star clusters and bona-fide galaxies. The objects
initially appeared in two main groups. On the lower mass end (M$_* \ga$ 10$^{6}$ -- 10$^{8}$ M$_{\sun}$) 
ultracompact dwarfs \citep[UCDs;][]{Hilker99,Drinkwater00}  emerged from the classical
GC population, while on the higher mass end ($\ga$ 10$^{9}$ M$_{\sun}$) the hitherto apparently rare compact 
ellipticals (cEs) were found to be relatively common \citep[e.g.][]{Mieske05,Chilingarian07,Chilingarian09,SmithCastelli08,Price09,AIMSSI,Chilingarian15}.
More recently the gap between star clusters and galaxies has finally been completely filled by systematic 
searches for CSSs, such as the first study in this series [\emph{A}rchive of \emph{I}ntermediate \emph{M}ass
\emph{S}tellar \emph{S}ystems (AIMSS) Survey; \citealt{AIMSSI},  \citetalias{AIMSSI} hereafter].

The final closing of the gap between star clusters and massive galaxies called into question the exact separation 
to be used to divide them, and even whether there was a fundamental difference between them at all
\citep[e.g.][]{ForbesKroupa11,Forbes11,Willman12}. In parallel to this debate, the emergence of an apparently 
tight scaling relation of size as function of stellar mass gave rise to the enticing prospect that it might be possible 
to unify all dynamically hot stellar systems from GCs to galaxies \citep[e.g.][]{Kissler-Patig06,Misgeld11} within
a single formation scenario. However, the subsequent discovery of additional objects which broadened the 
distributions of CSSs challenged this idea \citep[see e.g.][]{Brodie11,Forbes13}. Despite the confusion caused
by the burst of newly discovered CSS types, one recurring theme for CSSs over the whole mass range from UCDs to cEs is the 
suggestion that stripping processes play a role in their formation \citep[e.g.][]{Bassino94,Bekki01,Choi02,Drinkwater03,Bekki03}.

Today, there is little doubt that the tidal stripping of galaxies leads to the formation of many CSSs. The evidence
for this is compelling and varied, from CSSs caught in the act of formation and still embedded in tidal streams of stars from 
their disrupted progenitors \citep{Huxor13,Foster14,Jennings15}, to individual CSSs which host central 
supermassive black holes with masses expected to be found only in much more massive galaxies \citep{Kormendy97,Seth14}.
Furthermore, CSSs display stellar populations more akin to those of significantly more massive galaxies than to those of 
galaxies of similar mass  \citep[e.g.][]{Chilingarian09,Francis12,Sandoval15}. Finally there is the example of NGC~4546-UCD1
which is found to have a star formation history (SFH) which extends over several Gyr, a feat unlikely for a star cluster \citep{Norris15}.

Complicating this picture is the fact that there is also growing evidence that on either end of the CSS mass distribution 
many objects are simply continuations of the adjacent populations. Extrapolation of the GC luminosity functions 
of galaxies indicates that many if not most UCDs are simply GCs more massive than those found around the Milky 
Way, and which are only found in galaxies with sufficiently rich GC populations \citep[e.g.][]{Fellhauer02,Fellhauer05,Hilker06,Norris&Kannappan11,Mieske12}. 
Likewise there are suggestions that some cE galaxies may comprise the low mass tail of the true 
elliptical galaxy population (see e.g.~\citealt{Kormendy09}), rather than the end result of the tidal stripping of larger galaxies.
Counterintuitively, a further indication that the mechanisms responsible for forming CSSs may in fact 
be varied comes from the observation that they are found to be relatively ubiquitous. Both UCDs and cEs can be located
in all galactic environments from field (where tidal stripping is unlikely to be responsible) to 
dense galaxy clusters \citep{Norris&Kannappan11,Huxor13,Paudel14,AIMSSI,Chilingarian15}. As some CSSs are known to form
by stripping, and still others are found in environments where stripping is currently impossible, it seems to suggest that at least
one other formation mechanism is at work (or those are  run-aways that were stripped in a cluster and have been ejected via three body interaction,
see \citealt{Chilingarian15}).

Given the on-going discussion, new discoveries, and recent success in obtaining sizeable samples, the time is now ripe
to reexamine the information provided by the stellar populations of CSSs.
What can the stellar populations of CSSs reveal about their formation history, and can they be used as 
a discriminant between stripped objects and those that formed via other mechanisms?

In this paper, we report on the analysis of the integrated 
stellar populations. The sample is introduced in the following Section~\ref{section:sample}, and the
observations are described in Section~\ref{section:observations}. 
Section~\ref{section:stellpop} details the procedure
followed to obtain ages, metallicities, and  [$\alpha$/Fe] for all 
objects, while the results thereof are presented in Section~\ref{section:results}.
The results are further discussed in Section~\ref{section:discussion} and we conclude
with a summary of our findings in the final Section~\ref{section:summary}.

\section{Sample}
\label{section:sample}

  \begin{figure*} 
     \centering
   \includegraphics[scale=0.75,angle=-90]{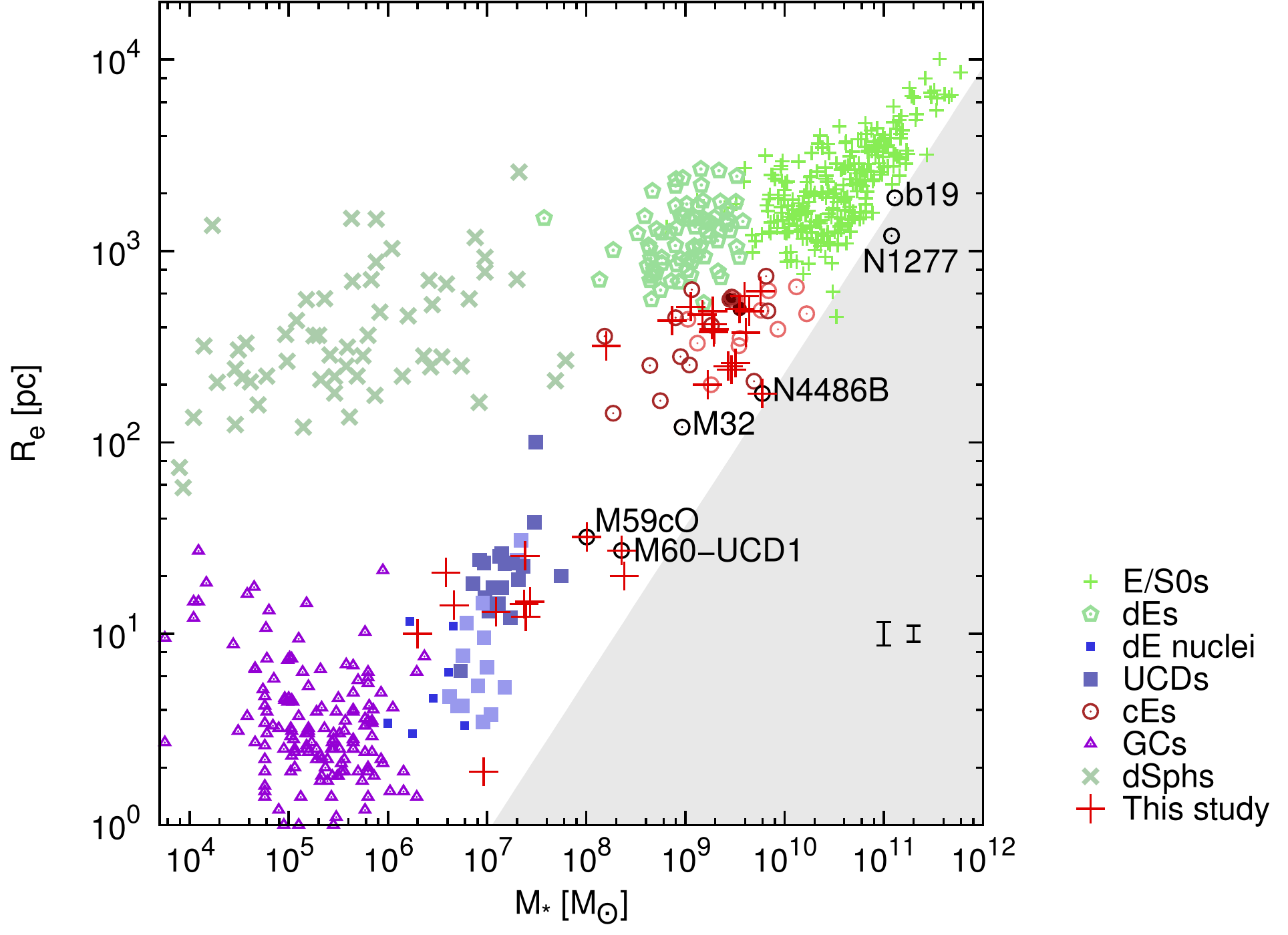}
   \caption{Size versus stellar mass plot for our sample objects, as well as other compact stellar systems from the literature, 
  and several comparison samples of various types of objects. 
   CSSs without a measurement of  [$\alpha$/Fe] are plotted with lighter colour.  Two isolated compact ellipticals \citep[cEs;][]{Huxor13,Paudel14} are highlighted with  filled dark red symbols. 
   Several objects that deserve special attention  are named in the plot (see text for details; b19 is short for SDSS J151741.75$-$004217.6). 
   Typical uncertainties for the sizes are shown in the lower right corner: the smaller bar for most of the CSSs and the larger for those within extended light, e.g.~within streams, and dSphs.
   The grey shaded area indicates the zone of avoidance, which \citet{Misgeld&Hilker11} found to be devoid of stellar systems.
   }
   \label{fig:massrad}
\end{figure*}

Our aim is to construct the largest possible sample of CSSs and comparison objects (dEs, dSphs, and E/S0s) 
with accurate spectroscopically-derived stellar population parameters. We therefore combine a large catalogue 
of literature data for CSSs and associated objects with new spectroscopic observations obtained with the 
Large Binocular Telescope (LBT), Keck II and Gemini South telescopes, targeted to fill in regions of parameter space which 
were previously undersampled.

We draw our sample for new spectroscopic observation principally from two main sources. The first is from 
Paper I  of the AIMSS project \citep{AIMSSI}, and the second is a new catalogue of cE galaxies 
selected from the SDSS (Huxor et al. in prep.). To increase the sample size further, and to broaden 
the range of parameter space studied, we have also observed additional literature CSSs which previously 
lacked suitable spectroscopically derived stellar population parameters. 
Furthermore, we include M60-UCD2 and VCC~165cE. 
These two objects were identified through using SDSS to select for UCD candidates based on colour, magnitude, and apparent size \citep[see][]{Sandoval15}.  M60-UCD2 was subsequently reported as a UCD candidate in the Next Generation Virgo Cluster Survey \citep[NGVS-J124352.42+112534.2][]{Liu15a}, and we adopt their photometric parameters for this object.

Note that our LBT/MODS spectroscopic 
observations were generally carried out as filler and bad weather programmes, and therefore
tends to preferentially focus on brighter, easier to observe targets. An overview of our full sample together 
with the comparison objects  is shown in the mass-stellar size plane in Fig.~\ref{fig:massrad}.

When selecting the literature comparison sample, we aimed at broadly sampling the population of dynamically 
hot stellar systems from the most massive early-type galaxies to CSSs from GCs to cEs. We required that the 
objects had spectroscopically determined stellar populations measurements and, in regions of parameter space 
for which there are many studies, we restricted the selection to homogenous sources with large samples. Our 
sources for the literature data are:

\begin{itemize}
\item GCs -- the Harris Milky Way GC catalogue \citep[][2010 edition]{Harris96}, with ages for 55 MW GCs 
 from \cite{Vandenberg13}, and  [$\alpha$/Fe] for 43 GCs from the compilation in \citet{Pritzl05}.
 \item UCDs - \citet{Paudel10b,Chiboucas11,Chilingarian11,Francis12}, with the photometry for some
 of the objects taken from \citet{Hasegan05,Evstigneeva07,Mieske08b,Zhang15}.
\item cEs -- \citet{Chilingarian07,Chilingarian09,Price09,Huxor11b,Huxor13,Paudel14,Guerou15}.
\item dSphs -- structural parameters and metallicities for Local Group dwarf spheroidal data are from \citet{McConnachie12}. Unfortunately,
no luminosity weighted age or [$\alpha$/Fe] information exists for these objects due to their low surface brightness. We include mass weighted ages
for some objects from \citet{Orban} as upper limits for the luminosity weighted ages.
\item dEs -- \citet{Chilingarian9}, \citet[main body of the dE after subtraction of nucleus]{Paudel11},  and \citet{Toloba2014b} with photometry and sizes from \citet{Janz08,Janz09}
\item dE nuclei -- \citet{Paudel11} with sizes from \citet{Cote06}
\item E/S0s -- the ATLAS$^{\rm 3D}$ survey  \citep{ATLAS3DI} with stellar population parameters from \citet{ATLAS3DXXX}, and using the multi-Gaussian expansions (MGE) of \citet{ATLAS3DXXI} to estimate the stellar mass within $R_\textrm{e}/8$.
\end{itemize}

Where possible we convert iron abundances to total metallicities using $[Z/\textrm{H}]=[\textrm{Fe}/\textrm{H}]+0.94 [\alpha/\textrm{Fe}]$
\citep{Thomas03}, and where this is not possible we indicate objects without  [$\alpha$/Fe] measurements, as for these objects the 
metallicity value is an approximation and is too low if the object is $\alpha$-enhanced.

\begin{table*}
\begin{center}
\caption{Observing Log}
\begin{tabular}{llllll} \hline
Name					&	R.A		& Dec.			&     Date   	&Telescope		& Setup												\\
						&	(J2000)	& (J2000)			& (dd/mm/yy)		& /Instrument		&													\\
\hline
2MASX J01491447+1301548                  		& 01:49:14.45  	& +13:01:55.1  		&  28/10/14   	& Keck / ESI         	&  0.75\arcsec\, 2400s 1.1\AA\, 0.8\arcsec \\
NGC 1128cE     	& 02:57:44.50  		& +06:02:02.2  		&  28/10/14  	&  Keck / ESI         	&  0.75\arcsec\, 1800s 1.1\AA\, 0.9\arcsec \\
	  					& 		  	& 				& 05/10/13 - 15/02/14   			&   LBT / MODS       	&  0.8\arcsec\, 9246s 2.3\AA\, 1.3 \arcsec \\
NGC~1272cE                  		& 03:19:23.04 	& +41:29:28.2  		&    28/10/14    	& Keck / ESI		&  0.75\arcsec\, 2400s 1.1\AA\, 0.8\arcsec \\
SDSS J075140.40+501102.6                  		& 07:51:40.39  	& +50:11:02.6  		& 15/02/14 	& LBT / MODS         	&  0.8\arcsec\, 7200s 2.3\AA\, 1.5\arcsec \\
                                   		& 		  	& 		  		& 28/10/14 	& Keck / ESI         	&  0.75\arcsec\, 1200s 1.1\AA\, 0.6\arcsec \\
NGC 2832cE			   	& 09:19:47.90  	& +33:46:04.9  		& 21/02/14   	&  LBT / MODS       	&  0.8\arcsec\, 1800s 2.3\AA\, 1.2 \arcsec \\
                                    		& 		  	& 		  		& 21/02/14   	&  Keck / ESI       	&  0.75\arcsec\, 3600s 1.1\AA\, 0.7\arcsec \\
NGC 2892cE                 		& 09:32:53.90  	& +67:36:54.5  		& 16/02/14   	&  LBT / MODS       	&  0.8\arcsec\, 5400s 2.3\AA\, 1.0\arcsec \\
cE0                    		 	& 09:47:29.23	& +14:12:45.3  		& 16/02/14   	& LBT / MODS        	&  0.8\arcsec\, 2700s 2.3\AA\, 1.0\arcsec \\
CGCG 036-042 			          	& 10:08:10.32 	& +02:27:48.3  		& 21/02/14   	& LBT / MODS        	&  0.8\arcsec\, 1800s 2.3\AA\,1.2 \arcsec \\
cE1                     			& 11:04:04.40  	& +45:16:18.9  		& 16-17/02/14  	& LBT / MODS        	&  0.8\arcsec\, 5400s 2.3\AA\, 1.0\arcsec \\
NGC 3628-UCD1              		& 11:21:01.20  	& +13:36:29.3   	&  21/03/14 	&   Keck / ESI         	&  0.75\arcsec\, 3600s 1.1\AA\, 0.7\arcsec \\
			             		& 		 	& 			   	&  13-14/06/15 	& LBT / MODS        	&  0.80\arcsec\, 3600s 2.3\AA\, 1.2\arcsec \\
NGC 3923-UCD1              		& 11:51:04.10   	& $-$28:48:19.8     	&  30/04/11  	&   Gemini / GMOS  	&  0.5\arcsec\, 10800s 1.26\AA\, 0.9\arcsec \\
NGC 3923-UCD2              		& 11:50:55.90   	& $-$28:48:18.4     	&   30/04/11   	&  Gemini / GMOS   	&  0.5\arcsec\, 10800s 1.26\AA\, 0.9\arcsec \\
NGC 3923-UCD3              		& 11:51:05.20  	& $-$28:48:58.9     	&  30/04/11  	&   Gemini / GMOS  	&  0.5\arcsec\, 10800s 1.26\AA\, 0.9\arcsec \\
PGC~038205                   		& 12:04:28.97  	& +01:53:38.8  		& 20/02/14  	&  LBT / MODS        	&  0.8\arcsec\, 2400s 2.3\AA\, 1.2\arcsec \\
M85-HCC1                  		& 12:25:22.84  	& +18:10:53.6  		& 04-06/04/14  	&  LBT / MODS        	&  0.8\arcsec\, 5400s 2.3\AA\, 1.4\arcsec \\
NGC 4486B / VCC1297		& 12:30:31.97  	& +12:29:24.6     	& 10/03/15  	&   LBT / MODS       	&  0.8\arcsec\, 1800s 2.3\AA\, 0.9\arcsec \\
S999                      			& 12:30:45.91  	& +12:25:01.5     	& 21/03/14   		&   Keck / ESI          	&  0.75\arcsec\, 11400s 1.1\AA\, 0.6\arcsec \\
NGC 4546-UCD1              		& 12:35:28.70  	& $-$03:47:21.1     	& 02/07/13 - 07/01/14  			&  Gemini / GMOS   	&  0.5\arcsec\, 22200s 1.41\AA\, 0.7\arcsec \\
Sombrero-UCD1             		& 12:40:03.13  	& $-$11:40:04.3     	&  15/03/15  			&  LBT / MODS        	&  0.8\arcsec\, 2700s 2.3\AA\, 1.5\arcsec \\
M59cO                     			& 12:41:55.33  	& +11:40:03.7  		& 10/03/15     	&  LBT / MODS       	&  0.8\arcsec\, 3600s 2.3\AA\, 1.1\arcsec \\
M59-UCD3                  		& 12:42:11.05  	& +11:38:41.2  		& 17/02/14  	&  LBT / MODS        	&  0.8\arcsec\, 3000s 2.3\AA\, 1.1\arcsec \\
                                     		& 		  	& 		  		& 21/03/14  	&  Keck / ESI        	&  0.75\arcsec\, 1200s 1.1\AA\, 0.7\arcsec \\
M60-UCD1	& 12:43:36.00  	& +11:32:04.6     	& 06/04/14 $\&$ 10/03/15   			&  LBT / MODS        	&  0.8\arcsec\, 6300s 2.3\AA\, 1.5 $\&$ 0.8\arcsec \\
M60-UCD2	& 	12:43:52.41	  	& 	+11:25:34.2	  		&  15/03/15   	& LBT / MODS        	&  0.8\arcsec\, 3600s 2.3\AA\, 1.8\arcsec \\
SDSS J133842.45+311457.0                  		& 13:38:42.45 	& +31:14:57.1  		& 30/03/14   	&  LBT / MODS       	&  0.8\arcsec\, 2700s 2.3\AA\, 2.0\arcsec \\
NGC~5846cE          & 15:06:34.27  	& +01:33:31.6  		& 12/03/15 	& LBT / MODS         	&  0.8\arcsec\, 1800s 2.3\AA\, 1.5\arcsec \\
2MASX J16053723+1424418                  		& 16:05:37.21  	& +14:24:41.3  		&  21/03/14  	& Keck / ESI          	&  0.75\arcsec\, 1800s 1.1\AA\, 0.7\arcsec \\
cE2 & 23:15:12.62     & $-$01:14:58.3 & 14/06/15&  LBT / MODS & 0.8\arcsec\, 2700s 2.3\AA\, 1\arcsec  \\
J233829.31+270225.1                   & 23:38:29.31     & +27:02:25.1     & 28/10/14 &   Keck / ESI           &  0.75\arcsec\, 2400s 1.1\AA\, 0.7\arcsec \\
\hline
\end{tabular}
\end{center}
{Notes: The setup lists the slitwidth, exposure time, spectral
resolution (FWHM, measured at around 5000\AA) and seeing. M60-UCD1 is also known as NGC~4649-UCD1, and NGC~1128cE as NGC~1128-AIMSS2. \label{table:log}}
\end{table*}

\section{Observations and data reduction}
\label{section:observations}
\begin{figure} 
   \centering
   \includegraphics[scale=0.285,angle=0]{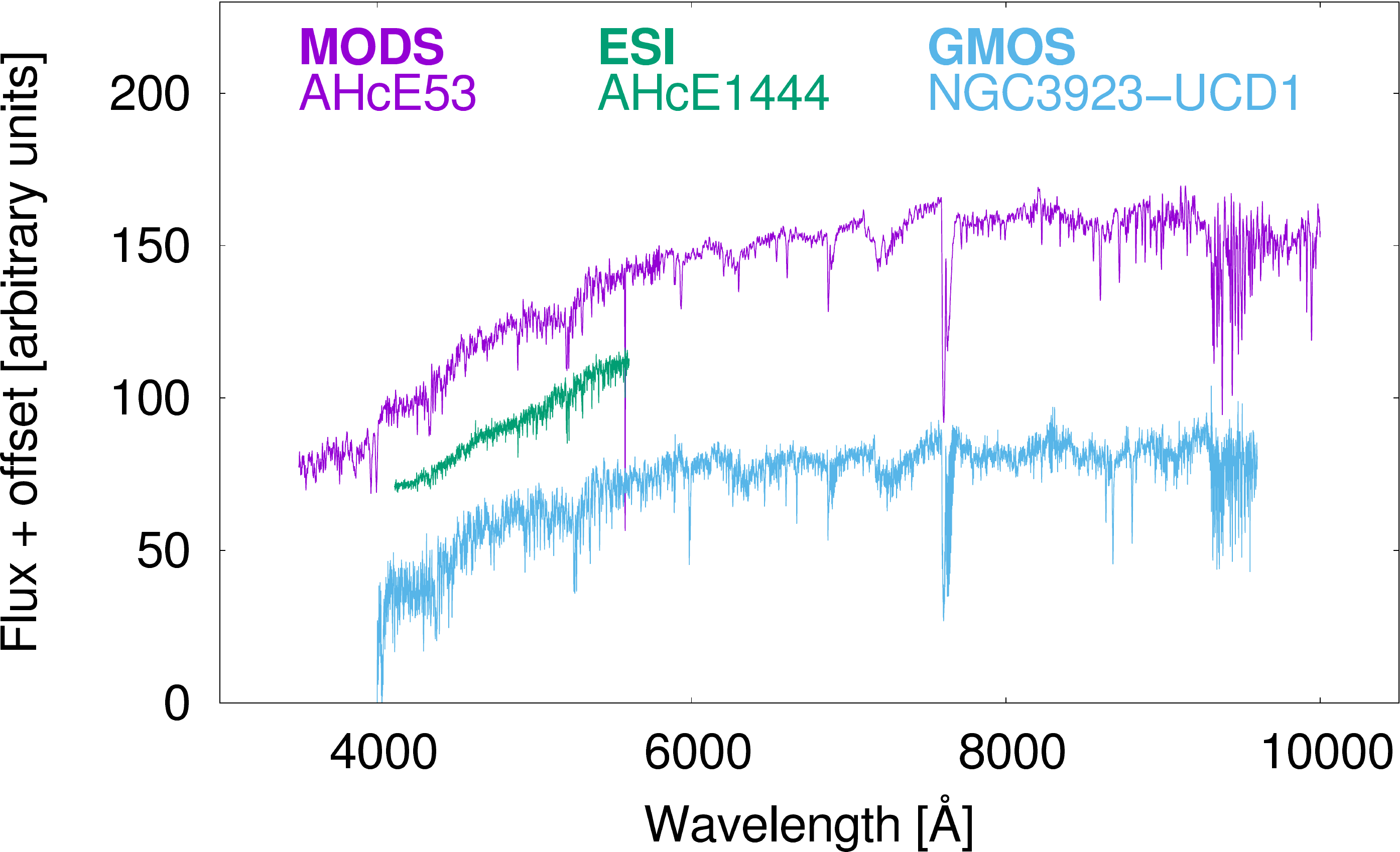}
   \caption{Example spectra observed with MODS, GMOS, and ESI, showing the different wavelength coverage. 
   Arbitrary scaling and offsets are applied to the fluxes for clarity. For MODS and ESI the very bluest $\sim$200\AA{} 
   are omitted, as they are of very low S/N and are not used in the analysis. At the red end the ESI spectrum is restricted to the wavelength range of 
   the comparison spectrum for joining the echelle orders (see text for details).  }
   \label{fig:spectra}
\end{figure}

In total to date we have obtained spectroscopy of 29 objects at the Large Binocular Telescope (LBT), Keck and Gemini 
observatories. Table \ref{table:log} provides the 
full observing log of the targets observed in this work, and in Fig.~\ref{fig:spectra} a sample spectrum for each telescope
is shown. 

\subsection{LBT MODS}

Observations of 19 of our CSSs were obtained with the first Multi-Object Double Spectrograph \citep[MODS1,][]{MODS} on the 
Large Binocular Telescope (LBT). MODS1 provides simultaneously observation of both a blue and red arm, providing
spectral coverage from $\sim$3200 to 10,000\AA\, split at around 5700\AA. For this study we primarily focus on the blue
arm which covers all of the main Lick absorption line indices and provides a spectral resolution of $\sim$2.3\AA\, FWHM, 
for our chosen slit width of 0.8\arcsec\, measured at around 5000\AA. Typical exposure times are of the
order of 3600s per target.

The MODS spectroscopy was reduced using the beta-release of the MODS reduction pipeline\footnote{The MODS reduction 
pipeline was developed by K.~Croxall with funding from NSF Grant AST-1108693. Details can be found at 
http://www.astronomy.ohio-state.edu/MODS/Software/modsIDL/}. This reduction comprised bias subtraction and
flat-fielding using python scripts which deal with the effects of MODS interlaced data readout, followed by 
wavelength calibration, object tracing and extraction, and finally flux calibration using observations of flux
standard stars observed during each observing run. We confirmed the reliability of the flux calibration procedure
by comparing the overlap region of the spectra produced by the reduction pipeline for the blue and red spectrographs.

\subsection{Gemini GMOS}
We incorporate very high quality spectra of four objects obtained with
Gemini Multi-Object Spectrograph \citep[GMOS,][]{Hook04} at Gemini-South.
The data for three NGC~3923 UCDs were reduced and first presented 
in  \citet{Norris12}, with the velocity dispersions measured  in 
the \citetalias{AIMSSI} paper. The observations were carried out in the
multi-object mode with the 1200 l/mm grating and 0.5\arcsec\, slitlets,
resulting in a spectral resolution of 1.26\AA, again measured around 5000\AA. For each of the three objects
there were 6 individual spectra with exposure times of 1800s each.
The spectrum for NGC~4546-UCD1 (\citealt{Norris15} and Escudero et al. in prep.)
was observed with essentially the same setup (with a spectral resolution of 1.41\AA{}) and reduced using the
same procedure.
The total integration time was 22200s split over 12 individual exposures.
In our chosen setup GMOS spectra cover a wavelength range from $\sim4100$ to 5600\AA{},
which means that one to three of the bluest Lick indices in the stellar population
analysis are missed (depending on the exact location of the slitlet on the mask).

\subsection{Keck ESI}

The spectra of 9 objects were taken with the Echellette Spectrograph and Imager \citep[ESI,][]{ESI} on the Keck II telescope.
The observations of each object were split into at least three individual exposures. The total integration times ranged from
1200 to 11400s. 
The instrument was used in the echellette mode with a slit width of 0.75\arcsec. The  wavelength range covered exceeds 4000 to 10000\AA{} across ten echelle orders, but the bluest and reddest parts are swamped in noise. 
For the stellar population analysis we concentrated on a region from 4050 to 5500\AA{} (in the restframe) with the Lick indices used for the model fitting.

The standard steps for the data reduction are conducted with \textsc{makee}.\footnote{Written by T.~Barlow, \url{http://www2.keck.hawaii.edu/inst/esi/makee.html}.}
The echelle characteristic of the spectrograph requires an additional step to bring the flux measurements in the different orders to a common level, and to join them for a uniform coverage of the whole spectral range. For that purpose we compared the spectrum of a star  observed under the same conditions to its reference spectrum (\citealt{STELIB}). 
The individual exposures and orders were combined in an $S/N$ optimized way with \textsc{uves\_popler}.\footnote{Written by M.~T.~Murphy, \url{http://astronomy.swin.edu.au/~mmurphy/UVES_popler/}.}  The individual errors of the spectral pixels were also propagated to produce a combined error spectrum.
The instrumental resolution of this setup is $\sim$1.1\AA{} (FWHM, measured around 5000 \AA{}) and thus higher than with MODS. The spectra were re-dispersed to 1\AA{} pix$^{-1}$ for the further analysis. 

~\\
We include the UCDs S999 \citep{Janz15} and NGC~3628-UCD1 \citep{Jennings15} in the sample, for which the same procedures were followed.
Five objects were observed twice with different instruments. We used these spectra to ensure that the stellar population analysis yields consistent results across the different observations (see Appendix). 
Likewise, we additionally analysed SDSS spectra of 4 objects for further comparison, 
and include results based on the SDSS spectrum of VCC~165cE (see Section~\ref{sec:VCC165cE}). 
For the subsequent analysis we use the error weighted averages of the stellar population parameters for the objects with multiple sets.

\subsection{Photometry}
Our sample of CSSs is based on the catalogue of \citetalias{AIMSSI} plus additional cEs selected from
the SDSS by Huxor et al. (in prep.). 
The photometry and size measurements were adopted from these studies
and we refer the reader to them for a detailed description of the analysis.
In summary \citetalias{AIMSSI} made use of \emph{HST} WFPC2, ACS or WFC3 imaging
to provide accurate size estimates and supplemented the available \emph{HST} imaging  (which was
generally only single or two band) with photometry from a variety of ground based sources
to provide wider wavelength coverage. 
Huxor et al.~(in prep.) use catalogued SDSS photometry except where the target
is judged to be deeply embedded in the halo of a larger galaxy. In this case 
the host galaxy light was subtracted following a scheme similar to that outlined in
\citetalias{AIMSSI}. Total magnitudes were then obtained with a curve-of-growth method 
and corrected for Galactic extinction following \citet{Schlafly11}.

For conversion to absolute magnitudes and physical scales, we used the distance
of the host galaxy, where applicable, or estimated the distance based on the object's recession
velocity assuming a Hubble flow with $H_0 = 68$ km s$^{-1}$ Mpc$^{-1}$.
The absolute magnitudes were converted to stellar mass using the mass-to-light ratios from \citet{Maraston05}
 and the stellar populations measured here, 
assuming a Kroupa initial stellar mass function. 
 Note that for high-mass UCDs the initial mass function (IMF) is debated \citep{Dabringhausen08,Mieske08a},  but the 
potential resulting shift in stellar mass does not change any of our conclusions ($\sim$0.27 dex in $\log M_*$ when changing from Kroupa to Salpeter IMF,
with nearly no difference for the model Lick indices, see \citealt{Maraston03}).
If the photometry was available in multiple filters, we used the reddest band when calculating
the stellar mass. A comparison to the previously used AIMSS stellar masses can be found in the Appendix.
For the literature samples we used the literature stellar masses when quoted, and followed the
same procedure otherwise.

\section{Analysis}
\label{section:stellpop}
All the reduced spectra were used as input to measure Lick line indices using the definitions of  \citet{Trager98}
with  \textsc{lector}.\footnote{Written by A.~Vazdekis, \url{http://www.iac.es/galeria/vazdekis/vazdekis_software.html}.} The measurements include 19 indices from H$\delta_\mathrm{A}$ to Fe5406.
With \textsc{pPXF} \citep{pPXF} the line-of-sight velocity distribution was fitted as well as a tenth order multiplicative polynomial to adjust the shape of the continuum.
As templates we used the \textsc{elodie}  library of  stellar spectra \citep{Prugniel07}.
 The polynomial was used to test whether the analysis for spectra that were not properly flux calibrated (especially the ESI spectra) could be biased. The effect on the final stellar population parameters was found to be negligible.
A Monte Carlo run with 50 random realizations of each spectrum using the error spectrum was carried out to obtain the statistical uncertainties of the index measurements.

The measured Lick indices need to be compared to model predictions in order to obtain stellar population characteristics such as age, metallicity, and  [$\alpha$/Fe]. For that the Lick indices need to be measured at the same resolution as the models 
 and corrected for offsets caused by the line-of-sight velocity distributions. We used the best-fitting template and the corresponding spectrum broadened to the object's velocity dispersion to obtain the offsets. The corrections for the higher moments in the velocity distributions are small and thus neglected.
 Then we interpolated the Lick index predictions of the high-resolution (2.5\AA{} FWHM) single stellar population (SSP) models of \citet{TMJ11} to a fine grid in $\log$ age,  $[Z/\textrm{H}]$, and $[\alpha/\textrm{Fe}]$ (0.02 dex in each direction) and used $\chi^2$ minimization to find the best-fitting model.
For this process two different sets of indices were used: a simple set of indices similar to ATLAS$^{\rm 3D}$  (H$\,\beta$, Mg$b$, Fe5270, and Fe5335)
and the full set of indices. For the latter case an iterative $\sigma$ clipping was applied to remove outliers, but the final set of indices was required to contain at least 6 indices including one Balmer line, and  a magnesium or iron index. A comparison of the resulting stellar population parameters is given in the Appendix,
as well as a comparison with literature stellar population parameters, and shows generally very good agreement. The so obtained quantities are luminosity weighted.

Generally, the small angular scale of the objects and the seeing during the observations
mean that the spectra are integrated over large apertures, and the resulting stellar 
population parameters are (luminosity-weighted) averages representative for the objects as a whole.

\begin{figure*}
\includegraphics[height=0.95\textwidth,angle=-90]{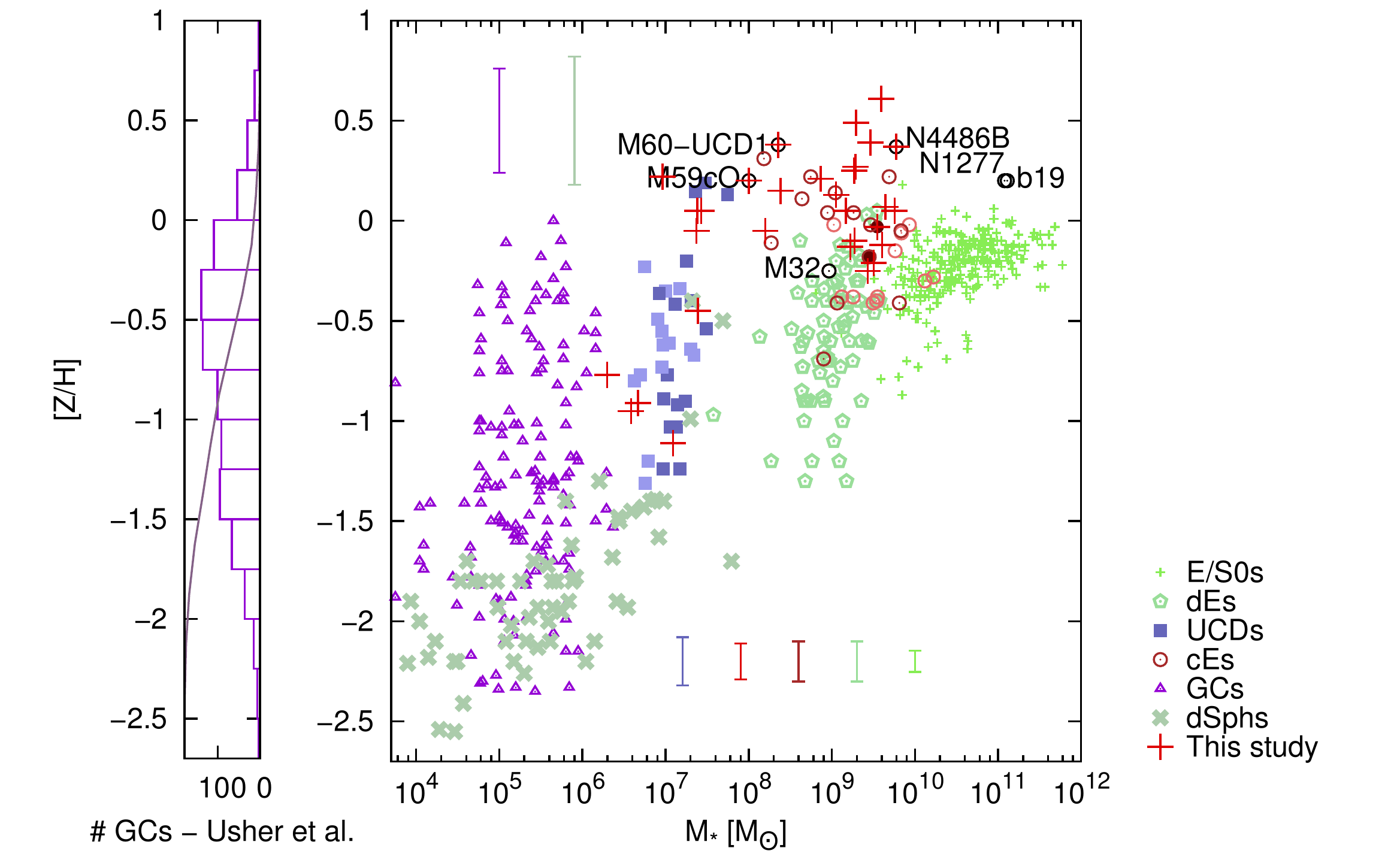}
\caption{Metallicity versus stellar mass for the same objects as in Fig.~\ref{fig:massrad}. 
Typical errorbars are given in the top and bottom parts of the main panel in the respective colours (for the GCs those of \citealt{Usher12} are shown).
Additionally, in the left panel the combined metallicity distribution of extragalactic globular clusters around various early-type galaxies from \citet{Usher12} is shown as a histogram, as well as  the cumulative distribution. The two isolated cEs of \citet{Huxor13} and \citet{Paudel14} are highlighted with filled dark red symbols.
   CSSs with lighter symbol colour are from the literature, but without any measurement of  [$\alpha$/Fe] (so their total metallicity is less secure).
The extremely high metallicity of most CSSs becomes evident, not only when compared to objects at the same stellar mass, but also generally. }
\label{fig:massmet}
\end{figure*}

\section{Results}
\label{section:results}
\begin{figure}
\centering
\includegraphics[height=0.48\textwidth,angle=-90]{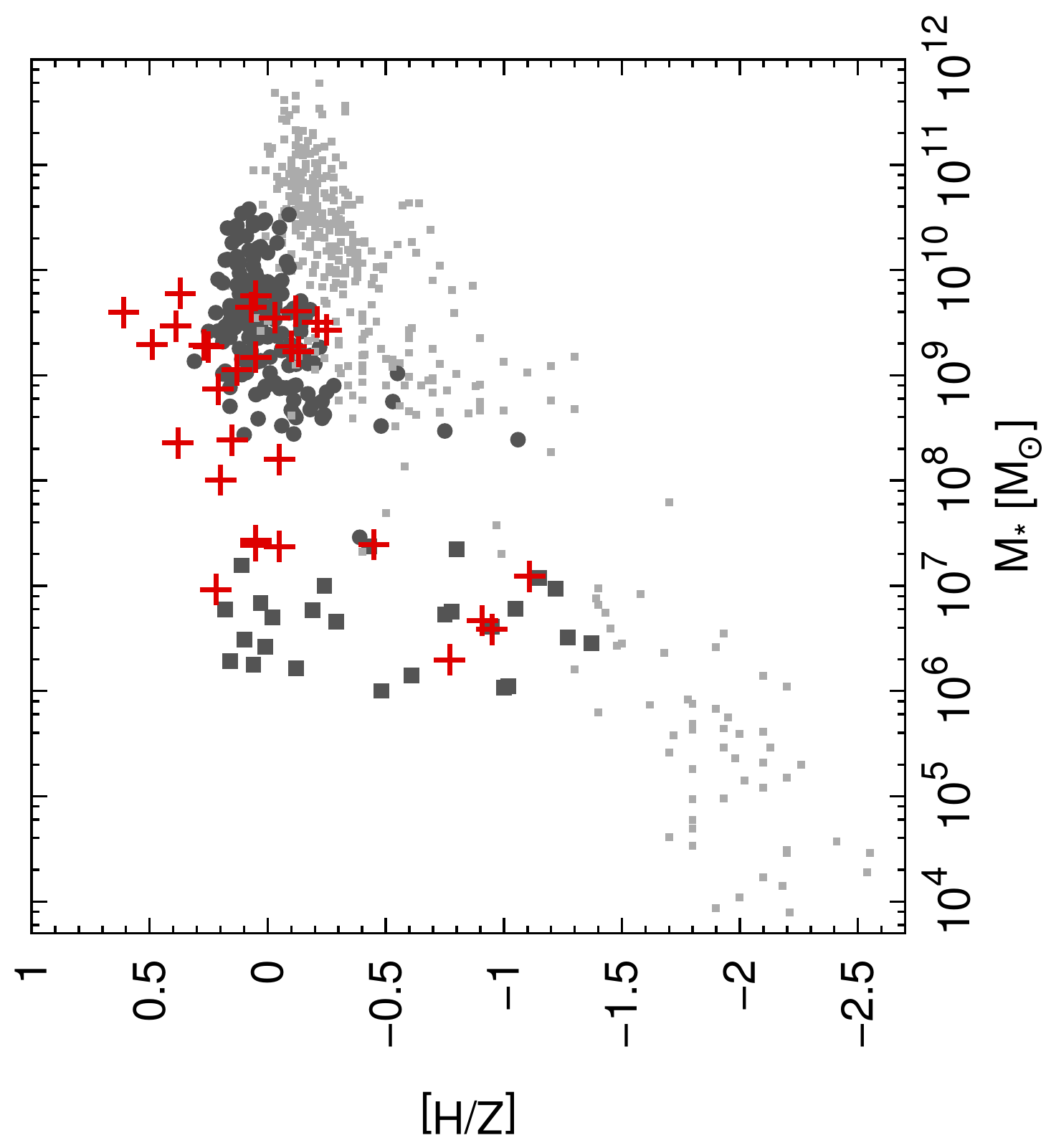}
\caption{Metallicity versus stellar mass  as in Fig.~\ref{fig:massrad}. 
Here the bona-fide galaxies (E/S0s, dEs, dSphs) are replotted with light grey symbols and our CSSs again with red crosses.
In addition, the ATLAS$^{\rm 3D}$ galaxies \citep{ATLAS3DXXX} are plotted a second time
with stellar mass and metallicity within $R_\textrm{e}/8$ (dark filled circles). The effect of metallicity gradients in massive early-type galaxies can be evaluated, and their inner parts can be compared to the CSSs.
The dark filled squares display the dwarf galaxy nuclei of \citet{Paudel11}.
The high metallicities of the CSSs are only found in the inner parts of other galaxies. }
\label{fig:innermassmet}
\end{figure}

The metallicities from our stellar population analysis are shown in Fig.~\ref{fig:massmet}. The comparison samples trace from dwarf spheroidals to giant ellipticals the well-known mass metallicity relation \citep[see e.g.][for large samples of massive early types from SDSS]{Gallazzi06,Thomas10} over many orders of magnitude in stellar mass, with only the most massive galaxies reaching solar metallicities. Our CSSs are almost exclusively more metal rich than the comparison galaxies. Especially when compared at constant stellar mass, they are clearly different.
While this applies also to the literature cEs, some of our CSSs are the most metal rich objects.
Most of the literature UCDs are less metal rich than our most extreme low-mass CSSs. Metallicities similar to those in the CSSs are basically only found in the inner regions
of galaxies (Fig.~\ref{fig:innermassmet}), such as the inner parts of the ATLAS$^{\rm 3D}$ galaxies \citep{ATLAS3DXXX}. At lower masses, the nuclei of dEs can have exceedingly high metallicities, when compared to the overall mass metallicity relation.

\begin{table*}
\caption{Basic and stellar population parameters of CSSs}
\begin{center}
\begin{tabular}{lccccccc}
\hline
Name & $M_* [\textrm{M}_{\sun}]$ & $R_\textrm{e} [\textrm{pc}]$ & $D$ [Mpc] & V [km s$^{-1}$] & Age [Gyr] &$[Z/\textrm{H}]$& $[\alpha/\textrm{Fe}]$ \\
\hline
NGC~4486B & $   6.0 \times 10^{9}$ &  180 &  16.4 &1509 & $14.5^{+0.4}_{-0.4}$ & $0.37\pm0.02$ &  $0.30\pm0.04$ \\
PGC~038205$^a$ & $   5.7 \times 10^{9}$ &  616 &  76.6 & 6335 & $14.5^{+0.9}_{-0.9}$ & $0.05\pm0.03$ &  $0.50\pm0.03$ \\
NGC~1128cE & $   4.4 \times 10^{9}$ &  484 & 100.0 &  7603& $ 8.5^{+2.2}_{-1.7}$ & $0.07\pm0.07$ &  $0.34\pm0.09$ \\
NGC~2832cE & $   4.1 \times 10^{9}$ &  375 &  98.6 &6841  &$14.3^{+2.1}_{-1.8}$ & $-0.12\pm0.20$ &  $0.48\pm0.17$ \\
NGC~2892cE & $   3.9 \times 10^{9}$ &  580 &  97.7 &6802  &$ 2.6^{+0.8}_{-0.6}$ & $0.61\pm0.09$ &  $0.50\pm0.06$ \\
{cE0} & $ 3.5 \times 10^{9}$&  499 &  85.5 & 5844 & {$ 6.0^{+1.9}_{-1.4}$} & {$-0.03\pm0.11$} &  {$0.25\pm0.11$} \\
cE2 &$   3.2 \times 10^{9}$ &  260 & 108.1 & 7580 &{$9.3^{+5.4}_{-3.4}$} & {$-0.21\pm0.20$} & $0.25\pm0.10$ \\
NGC~5846cE & $   2.9 \times 10^{9}$ &  240 &  26.7 & 1479 & $14.5^{+1.4}_{-1.3}$ & $0.39\pm0.03$ &  $0.34\pm0.04$ \\
J233829.31+270225.1 & $   2.7 \times 10^{9}$ &  250 & 134.3 &9968 & $12.6^{+2.8}_{-2.3}$ & $-0.25\pm0.07$ &  $0.38\pm0.12$ \\
cE1 & $   2.0 \times 10^{9}$ &  390 &  92.0 & 6391 & $ 2.8^{+0.4}_{-0.3}$ & $0.49\pm0.06$ &  $0.16\pm0.04$ \\
SDSS J075140.40+501102.6 & $   1.9 \times 10^{9}$ &  485 &  87.1 & 6174 & $ 9.2^{+2.9}_{-2.2}$ & $-0.10\pm0.10$ &  $0.49\pm0.12$ \\
2MASX J01491447+1301548 & $   1.9 \times 10^{9}$ &  414 &  71.1 & 4861& $ 7.9^{+1.4}_{-1.2}$ & $0.25\pm0.06$ &  $0.12\pm0.03$ \\
NGC~1272cE & $   1.9 \times 10^{9}$ &  377 &  76.2 &3693  &$ 9.5^{+1.2}_{-1.1}$ & $0.27\pm0.05$ &  $0.32\pm0.04$ \\
{VCC~165cE} & $   1.7 \times 10^{9}$ &  200 & 180.0  & 12694 & {$1.5^{+0.4}_{-0.3}$} & {$-0.13\pm0.16$} & {$0.20\pm0.12$} \\
CGCG 036-042 & $   1.5 \times 10^{9}$ &  465 &  32.5 & 2062 & $10.0^{+2.2}_{-1.8}$ & $0.05\pm0.08$ &  $0.22\pm0.05$ \\
2MASX J16053723+1424418 & $   1.1 \times 10^{9}$ &  511 &  67.9 & 4833 & $ 2.4^{+0.2}_{-0.2}$ & $0.13\pm0.05$ &  $0.06\pm0.04$ \\
SDSS J133842.45+311457.0 & $   7.4 \times 10^{8}$ &  433 &  71.1 &4604  &$ 4.6^{+1.8}_{-1.3}$ & $0.21\pm0.12$ &  $0.12\pm0.06$ \\
M59-UCD3 & $   2.4 \times 10^{8}$ &   20 &  14.9 & 429 &$11.7^{+3.0}_{-2.4}$ & $0.15\pm0.10$ &  $0.29\pm0.05$ \\
{M60-UCD2} & $   2.4 \times 10^{7}$ &   14 &   16.4 & 791  &{$7.6^{+3.9}_{-2.6}$} & {$-0.05\pm0.13$} & {$0.18\pm0.10$} \\
M60-UCD1 & $   2.3 \times 10^{8}$ &   27 &  16.4 & 1278 &$14.5^{+3.7}_{-3.0}$ & $0.38\pm0.07$ &  $0.33\pm0.04$ \\
M59cO & $   1.0 \times 10^{8}$ &   32 &  14.9 &  723 &$14.5^{+4.6}_{-3.5}$ & $0.20\pm0.20$ &  $0.26\pm0.10$ \\
Sombrero-UCD1 & $   2.7 \times 10^{7}$ &   14 &   9.0 &1327 & $14.5^{+1.7}_{-1.5}$ & $0.05\pm0.05$ &  $0.22\pm0.04$ \\
NGC~3923-UCD1 & $   2.5 \times 10^{7}$ &   12 &  21.3 &2135 & $ 8.3^{+1.9}_{-1.6}$ & $-0.45\pm0.09$ &  $-0.04\pm0.06$ \\
NGC~4546-UCD1 & $   2.4 \times 10^{7}$ &   25 &  13.1 & 1210& $ 5.8^{+0.1}_{-0.1}$ & $0.05\pm0.01$ &  $-0.04\pm0.02$ \\
NGC~3923-UCD2 & $   1.2 \times 10^{7}$ &   13 &  21.3 &1494 & $ 9.5^{+2.4}_{-1.9}$ & $-1.11\pm0.14$ &  $-0.30\pm0.19$ \\
M85-HCC1 & $   9.2 \times 10^{6}$ &    1.9 &  17.9 & 699 &$ 1.9^{+1.4}_{-0.8}$ & $0.22\pm0.15$ &  $0.05\pm0.11$ \\
NGC~3923-UCD3 & $   4.6 \times 10^{6}$ &   14 &  21.3 &2322 & $ 8.3^{+3.3}_{-2.4}$ & $-0.91\pm0.16$ &  $0.30\pm0.20$ \\
S999 & $   3.8 \times 10^{6}$ &   20 &  16.8 & 1504& $ 7.6^{+2.0}_{-1.6}$ & $-0.95\pm0.12$ &  $0.34\pm0.10$ \\
NGC~3628-UCD1 & $   2.0 \times 10^{6}$ &   10 &  10.6 & 824& $ 6.6^{+1.4}_{-1.2}$ & $-0.77\pm0.16$ &  $-0.08\pm0.15$ \\
\hline\end{tabular}
\end{center}
Notes:  The distance gives the distance assumed in this work, which is in some cases based on the Hubble flow due to the lack of direct measurements ($H_0 = {68}$ km s$^{-1}$ Mpc$^{-1}$). The uncertainties of the recession velocities estimated from the Monte Carlo simulations are smaller than 10 km s$^{-1}$ so that they are dominated by the systematics. The stellar population parameters are SSP equivalents. The uncertainties in the stellar population parameters are from the Monte Carlo simulations of the whole index measurement and $\chi^2$-minimization process based on the error spectra (see text).    $^a$For this object the GALFIT measurement of the size failed, so aperture photometry was done instead. The value for the radius is to be taken with caution due to the bright halo of the host galaxy.
\label{table:pars}
\end{table*}

\begin{figure*}
\includegraphics[height=0.51\textwidth,angle=-90]{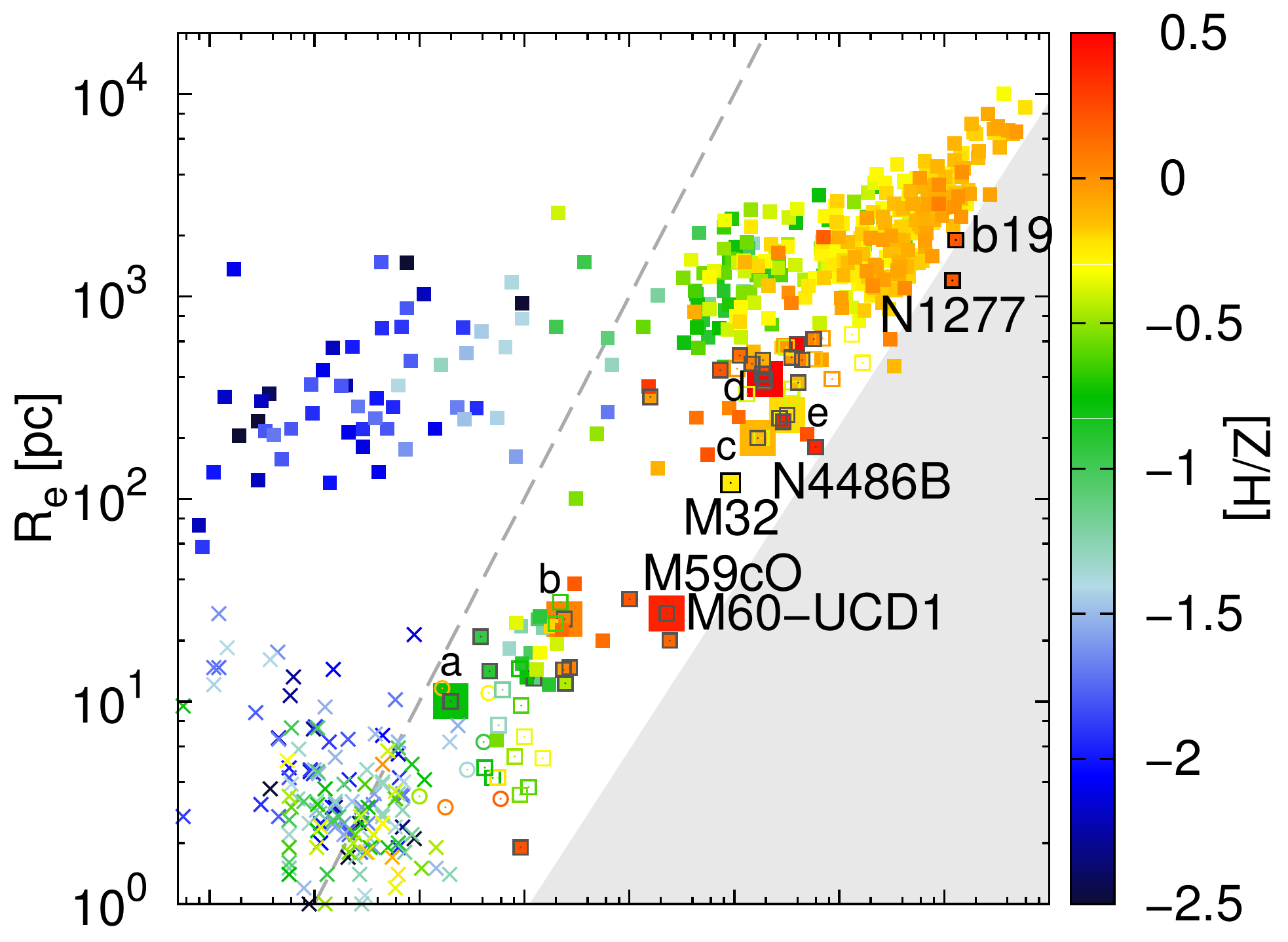}\ \ \ \includegraphics[height=0.4\textwidth,angle=-90]{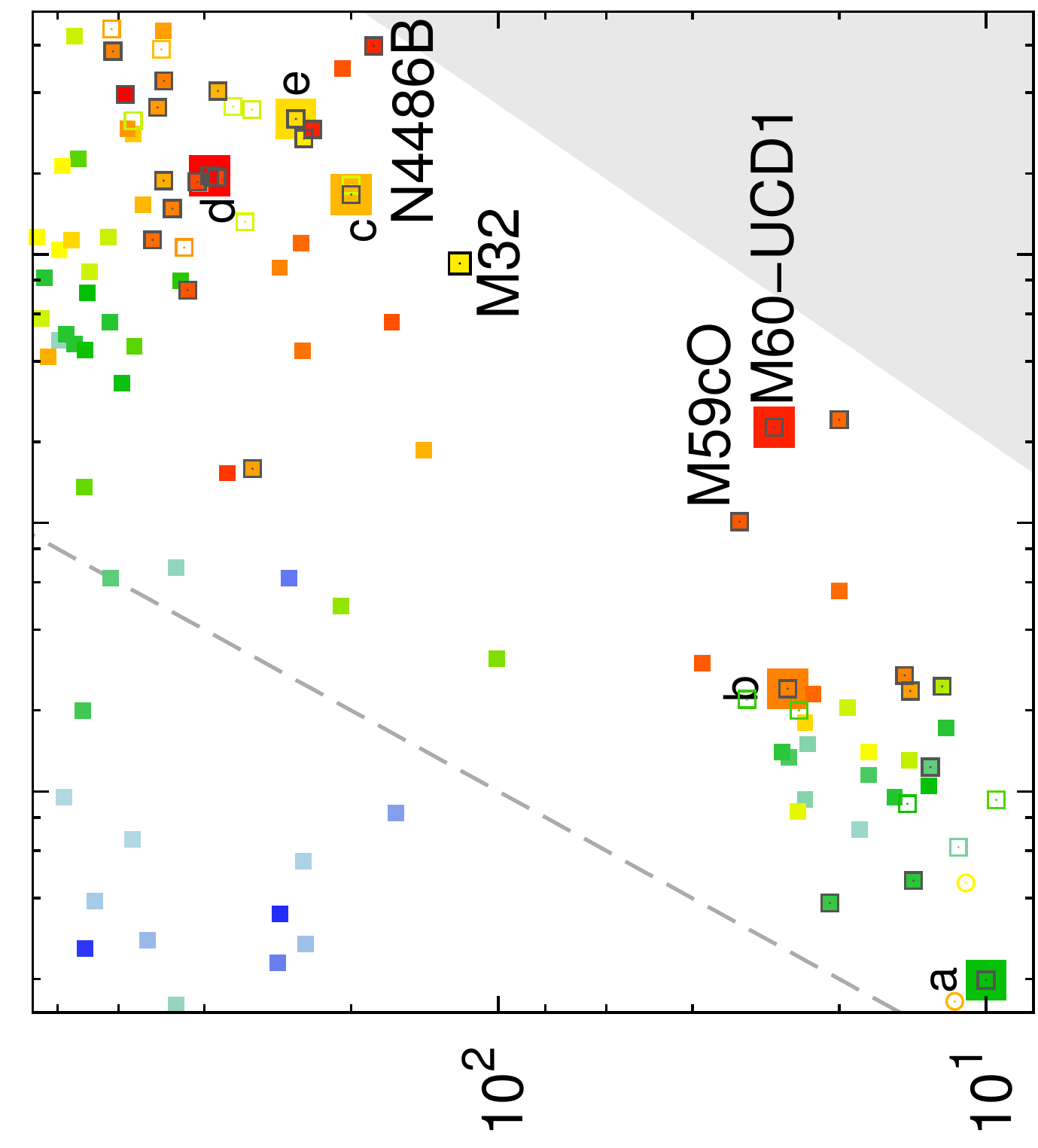}\\
\includegraphics[height=0.51\textwidth,angle=-90]{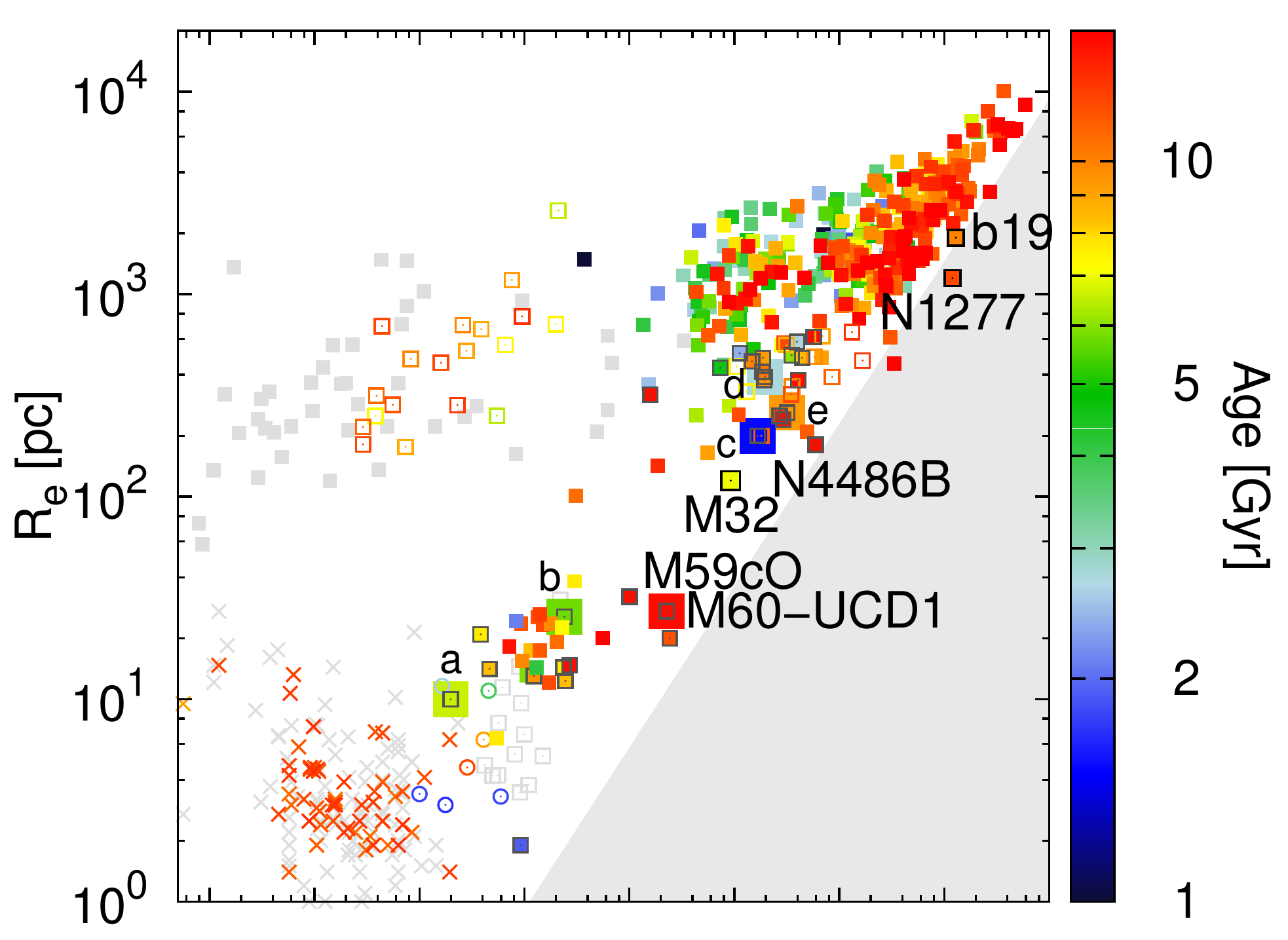}\ \ \ \includegraphics[height=0.4\textwidth,angle=-90]{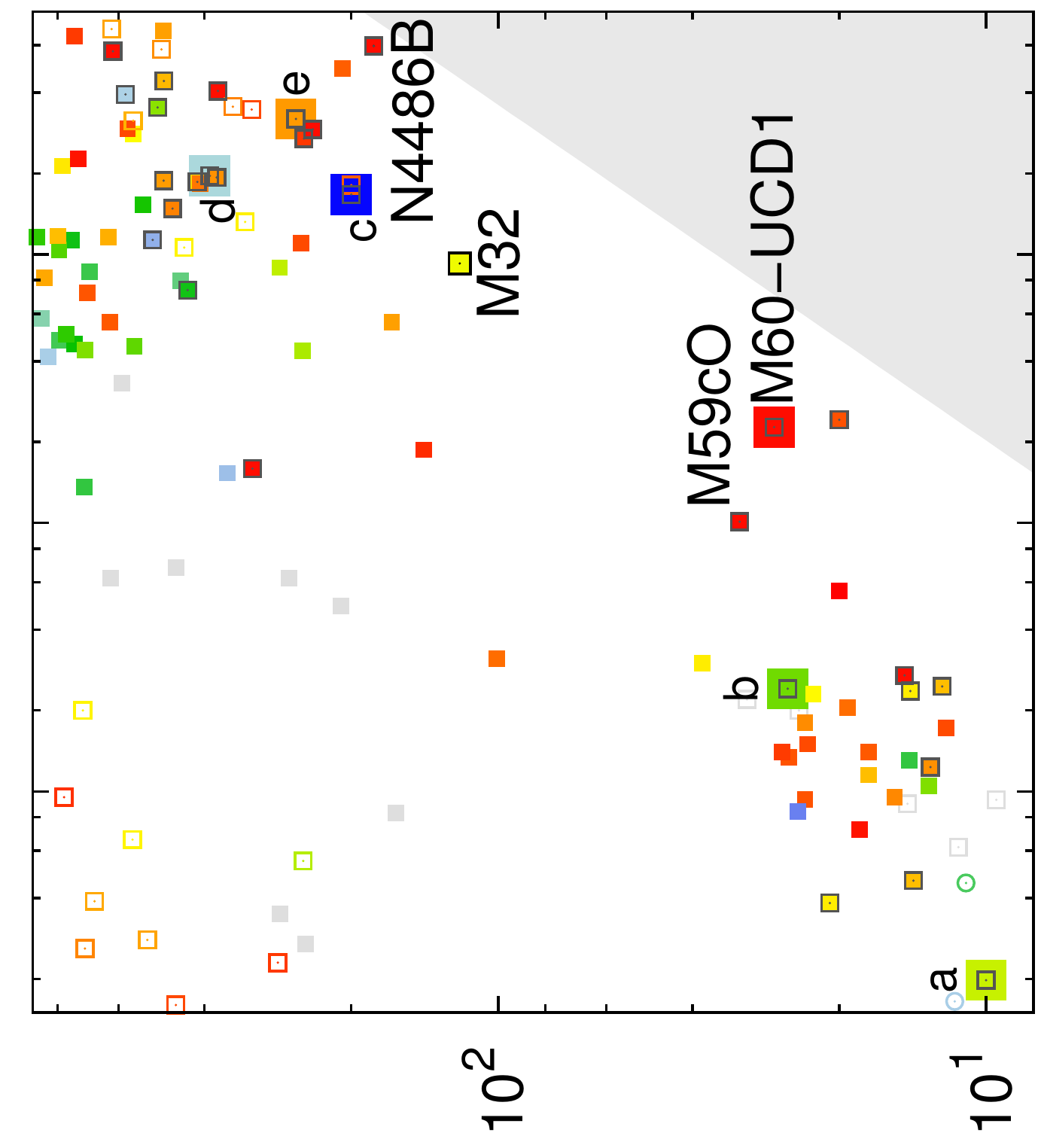}\\
\includegraphics[height=0.51\textwidth,angle=-90]{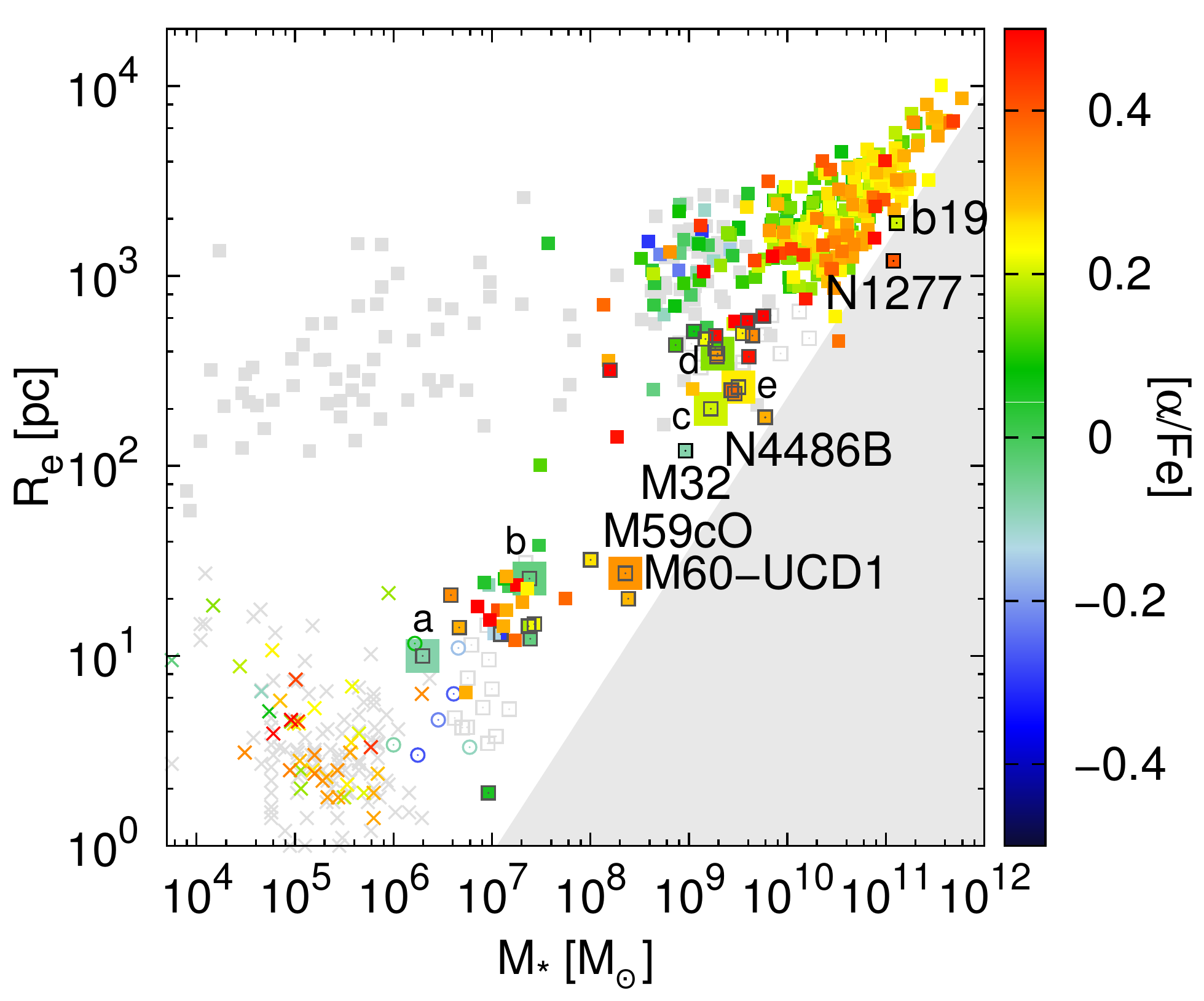}\ \ \ \includegraphics[height=0.4\textwidth,angle=-90]{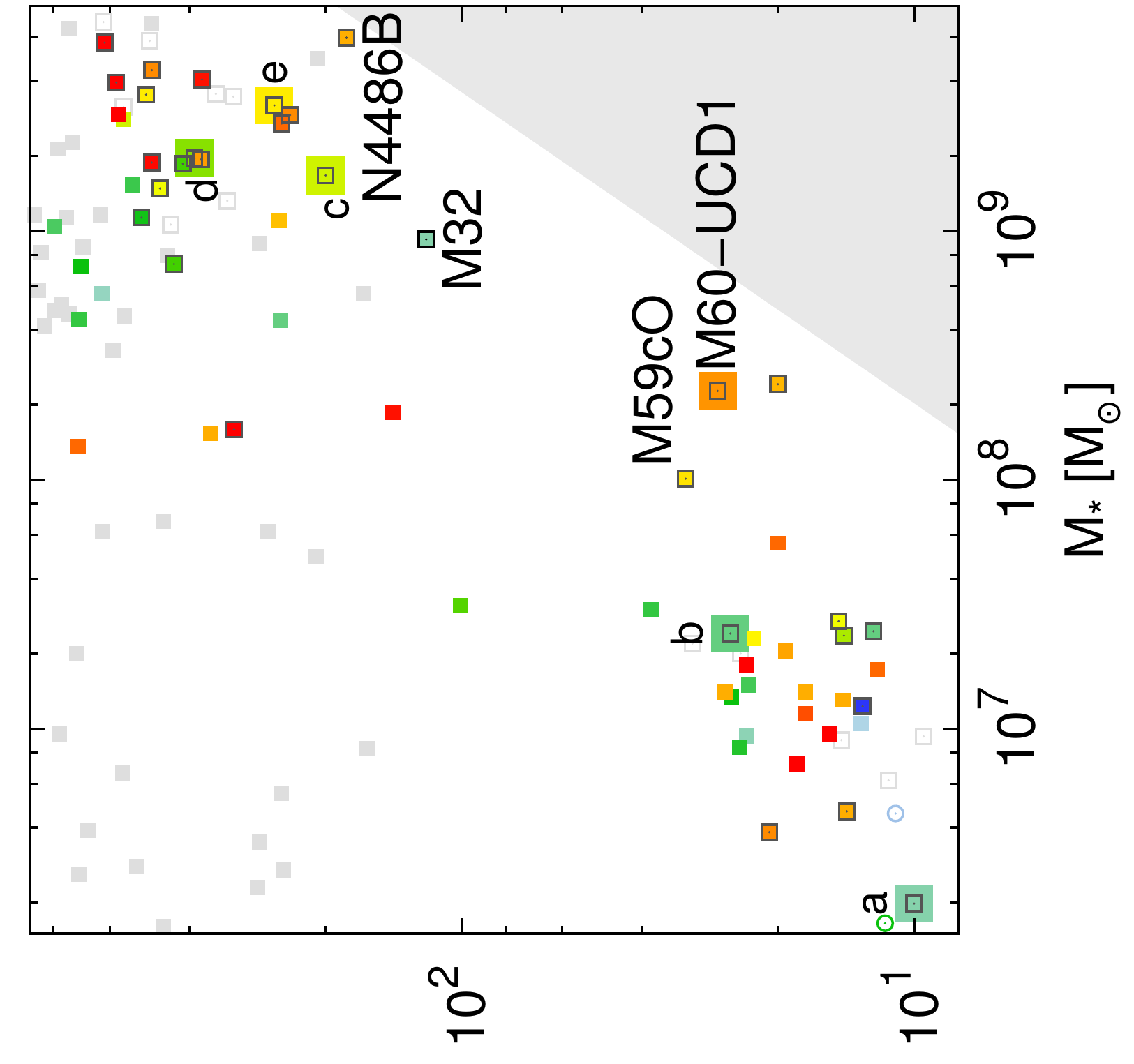}
\caption{Size--stellar mass plots colour coded by stellar population characteristics. The colours display from top to bottom the SSP-equivalent  metallicity, age, and  [$\alpha$/Fe]. 
The panels in the right column are zoom-ins to the parameter space of our CSSs.
Globular clusters are displayed with crosses, dwarf nuclei with open circles, and the rest with squares.  
Literature CSSs without a measurement of  [$\alpha$/Fe] are shown with open squares. The ages of dSphs are also plotted with open squared, since they are mass-weighted averages, which are upper limits for the luminosity weighted ages. Objects with  information lacking are plotted grey.
The CSSs (our objects are highlighted with grey borders) exhibit metallicities that exceed those of other objects at the same mass. At low mass a number of them separate from globular clusters by exhibiting younger ages. 
The objects discussed in Section \ref{section:stripped_objects} are highlighted with large symbols (in order of increasing stellar mass: NGC~3628-UCD1 -- a, NGC~4546-UCD1 -- b, M60-UCD1, VCC~165cE -- c, and  cE1 and cE2 from \citealt{Huxor11b} -- d,e). The grey dashed line in the middle panel is a line of constant velocity dispersion inferred from the virial theorem with constant virial coefficient. 
\label{fig:massrad_stellpop}}
\end{figure*}

The metallicities, as well as stellar ages and  [$\alpha$/Fe], are shown as parameters in the size stellar mass plane in Fig.~\ref{fig:massrad_stellpop}, with all parameters also being listed in Table \ref{table:pars}.
For the metallicities it can be seen again that the CSSs are more metal rich than more diffuse galaxies at the same mass. 
Another aspect becomes evident. Overall the metallicities seem to increase along lines  of increasing velocity dispersion, rather than stellar mass, as observed also for more massive galaxies \citep[e.g.][compare also to the surface densities in Fig.~\ref{fig:stellpop_add}]{ATLAS3DXXX,Guerou15}.
Turning to stellar ages, compact objects at the low- and high-mass end, i.e.\ GCs and giant early-type galaxies, are generally old. 
While this applies also to quite a number of CSSs,
many of those studied here exhibit  ages of $\sim$2-8 Gyr.
Finally, the CSSs show varying  levels of  [$\alpha$/Fe], with some of them reaching
 the  [$\alpha$/Fe] of massive galaxies, others being moderately enhanced similar
to GCs, and yet others having solar   [$\alpha$/Fe].

\subsection{CSSs likely formed by stripping}
\label{section:stripped_objects} 
There is good evidence that six of our CSSs were formed via stripping, and we discuss them below.
These objects (NGC~3628-UCD1 -- \emph{a},  NGC~4546-UCD1 -- \emph{b}, M60-UCD1,  VCC~165cE -- \emph{c}, cE1 -- \emph{d}, cE2 -- \emph{e}) are also highlighted in Fig.~\ref{fig:massrad_stellpop}, where
stellar age, metallicity, and  [$\alpha$/Fe] are shown as
a parameter in the size stellar mass plane.
Two of the six objects have  super solar metallicity,
and all of them are  at least 0.3 dex  more metal rich than more diffuse galaxies
at similar stellar mass. Two of the stripped high mass CSSs (cE1 and VCC~165cE)
show young ages ($<$3 Gyr).
At low stellar mass, two of the stripped CSSs (NGC~3628-UCD1 and NGC~4546-UCD1) 
are younger in comparison to the Galactic GCs.
 While most of the Galactic GCs are $\alpha$-enhanced, two of the three low-mass stripped CSSs (NGC~3628-UCD1 and NGC~4546-UCD1) 
 show  [$\alpha$/Fe] close to solar or below.

\subsubsection{NGC~3628-UCD1}
\citet{Jennings15} described NGC~3628-UCD1 (\emph{a} in Fig.~\ref{fig:massrad_stellpop}) as an $\omega$Cen-like
object caught in formation via tidal stripping, as it is found within a stellar stream
of material of a progenitor object that has been torn apart by  NGC~3628.
This object is one of those that were observed twice. Both the ESI and MODS spectra
result in very consistent age, metallicity, and  [$\alpha$/Fe].
The metallicity ($[Z/\textrm{H}]=-0.77$ dex) is moderate in comparison to other CSSs, and brings
NGC~3628-UCD1 closer to the galaxy mass--metallicity relation. Compared to GCs,
the intermediate age and (sub-) solar  [$\alpha$/Fe] set the object apart, consistent
with the formation via stripping instead of a GC-type object.

\subsubsection{NGC~4546-UCD1}
Using full spectral fitting of a high S/N spectrum of NGC~4546-UCD1 (\emph{b} in Fig.~\ref{fig:massrad_stellpop}) 
\citet[using the same spectrum as examined here]{Norris15} found that this object was actively forming stars from early epochs until quite recently.
This prolonged star formation is unexpected for a star cluster and, taken together with the observation
that this object counterrotates its host galaxy, indicates that the object was once the nucleus of a dwarf galaxy 
that was stripped by NGC~4546 relatively recently. Despite using the same spectrum and similar stellar population
models, the luminosity weighted age and metallicities we derive here for NGC~4546-UCD1 are slightly
inconsistent with those of \citeauthor{Norris15} (\citeyear{Norris15}; age of $5.8\pm0.1$ vs $4.0^{+0.93}_{-0.75}$ Gyr, $[Z/\textrm{H}]=0.05\pm0.01$ vs $0.18\pm0.06$ dex), 
difficulty of achieving consistent stellar population parameters when even slightly different analysis codes 
or models are used. Nevertheless our measurements confirm that NGC~4546-UCD1 is relatively young and metal rich, at
least compared to the vast majority of GCs, and furthermore has near solar  [$\alpha$/Fe].
 This last fact is expected given the observation that NGC~4546-UCD1 had an extended
SFH \citep{Norris15}.

\subsubsection{M60-UCD1}
\citet{Seth14} identified a supermassive central black hole in
M60-UCD1, which accounts for a  disproportionately high fraction of the
object's total mass. This is taken as evidence of originating from a more massive galaxy.
In contrast to the other objects in this category, M60-UCD1 has a very old age.
Like the overly massive black hole, the extremely high metallicity suggests that the object
was more massive in its past.
Our stellar population parameters are consistent with \citet{Strader13}, who reported 
solar  iron abundances, old age ($14.5\pm0.5$ Gyr), and  super solar $[\alpha/\textrm{Fe}]$$\sim$+0.2 dex.

\subsubsection{VCC~165cE}
\label{sec:VCC165cE}

 \citet{Paudel13} described a compact source close to VCC~165 as a background AGN. However,
 the current SDSS spectrum \citep{SDSSDR10} reveals that VCC~165 has a recession velocity exceeding 12000 km s$^{-1}$ and
 close to that of the compact object in its vicinity. While the old velocity measurement of 255 km s$^{-1}$ (NED) placed VCC~165
 in the Virgo cluster, like the NGC~4216 system analysed by \citet{Paudel13}, the new value puts it behind the Virgo cluster
 together with the compact object. The stream connecting the two objects is clearly visible in the NGVS image of \citet{Paudel13} and likely belongs 
 to this system in the background. Therefore, the compact object should be considered as a cE in formation, and we called it VCC~165cE. 
The stellar population parameters with a young age ($1.5^{+0.4}_{-0.3}$ Gyr), metallicity around solar ($[Z/{\rm H}]=-0.13\pm0.16$), 
and $[\alpha/\textrm{Fe}]=0.20\pm0.12$ fit very well to this scenario.

\subsubsection{cE1 and cE2}
  \citet{Huxor11b} discovered two cEs embedded in tidal streams.
For cE1  (\emph{d} in Fig.~\ref{fig:massrad_stellpop}) our results and their analysis based on the SDSS spectrum
agree qualitatively: young to intermediate age, supersolar metallicity, and
slightly enhanced  [$\alpha$/Fe]. Here we obtain a somewhat younger
age and higher metallicity. For the age of cE2  (\emph{e} in Fig.~\ref{fig:massrad_stellpop}),  the agreement is less good. 
The analysis of the MODS spectrum suggests that the object is dominated by old stars, unlike the intermediate
age of $5.4\pm1.6$ Gyr found by \citet{Huxor11b}. 
In order to further investigate the difference, we analysed also the SDSS spectrum. 
The SDSS spectrum has an $S/N$ of 8--15, while our MODS spectrum has 30--50.
For the SDSS spectrum the two different sets of indices (see Appendix) yield different ages. 
The full index set results in an age similar to the one obtained with the MODS spectrum.
Only when using the minimal set (H$\,\beta$, Mg$b$, Fe5270, and Fe5335)
do we also obtain an intermediate age.

\subsubsection{Other candidate stripped CSSs}
We note that  9 additional cEs were observed to be embedded in tidal streams \citep{Chilingarian09,Chilingarian15},
and should be considered as examples with strong direct evidence for a stripping origin.

We also note that the newly discovered object M59-UCD3 \citep{Sandoval15} probably fits in this category, since its measured properties
make it a virtual clone of M60-UCD1; its stellar mass (2.4 vs 2.3 $\times$10$^8$ M$_{\sun}$), size
(20 vs 27 pc), and most importantly its extremely high velocity dispersion ($\sim70$ km s$^{-1}$ from our MODS spectrum vs $68\pm5$ km s$^{-1}$; \citealt{Strader13}) mean that  this object is very likely to host a supermassive black hole like M60-UCD1. 
Our two spectra (from MODS and ESI) result in consistent stellar population parameters, which are also consistent
within the uncertainties with the analyses of \citet{Sandoval15} and \citet{Liu15b}.
The latter study reported a slightly higher velocity dispersion of $78$ km s$^{-1}$.
Similar to M60-UCD1, M59-UCD3 also has super solar metallicity and  [$\alpha$/Fe].

S999, while not featuring a very high metallicity, may also fall into the category of stripped objects. 
\citet{Hasegan05} found an extremely high mass-to-light ratio for S999,
which was recently confirmed by \citet{Janz15}. The latter study also analysed 
the stellar populations of S999, and argued that  the stars alone (in dynamical equilibrium) cannot account
for the mass-to-light ratio. The authors concluded that the apparently too high
dynamical mass probably relates to a tidal stripping event.
The moderate metallicity of S999 ($[Z/{\rm H}]=-0.95\pm0.12$) shows that excessive amounts of metals
may be seen as sufficient for suggesting stripping formation but not 
absolutely necessary.

\section{Discussion}
\label{section:discussion}

For CSSs of all masses it has been suggested that stripping of originally more massive galaxies
is (at least) one formation channel \citep[e.g.][]{Faber73,Bekki01,Pfeffer13,AIMSSI}.
The following observations are the key for this idea: CSSs are prevailingly found in
environments with high galaxy density and proximity to more massive galaxies, 
their velocity dispersions are more akin to more massive galaxies than to more diffuse galaxies at fixed stellar mass \citep[e.g.][]{Chilingarian09}. 
The latter also applies to the CSS metallicities, which exceed those of galaxies at the same stellar mass.

Here we confirm this for a sample of CSSs spanning a large range in stellar mass, and find generally high metallicities  (Fig.~\ref{fig:massmet}).
Indeed, the metallicities are \emph{too high} when compared to the mass metallicity relation, which is beautifully traced
over many orders of magnitudes in stellar mass by our comparison samples.
When comparing the metallicity values for the various objects, the details of the measurements have to be considered.
For our measurements the value is representative for the stellar population of the entire object, due to the
small angular scale and the seeing. The situation is definitely different for the large early-type galaxies,
as can be seen in Fig.~\ref{fig:massmet}, where the ATLAS$^{\textrm{3D}}$ are plotted twice.
The metallicity gradients make the inner parts substantially more metal rich than the stars in the outskirts. 
Indeed, the extremely high metallicities are only paralleled in the inner parts of more massive 
galaxies ($R_\textrm{e}/8$ and the mass within this radius of the ATLAS$^{\textrm{3D}}$ galaxies happen to
match those of the cEs quite closely). This can be confirmed by comparison to the mass metallicity relation of massive early-type galaxies
as traced by SDSS \citep{Gallazzi06,Thomas10}, which is shifted to higher metallicities in comparison to the
ATLAS$^{\textrm{3D}}$ relation. Those reach super solar metallicities, since the small fibre of the spectrograph selects the stars in the galaxies' central regions.
The mass metallicity relation of \citet{Thomas05} using the metallicities of early-type galaxies within $R_{\textrm{e}}/10$
passes through the  points for the inner parts of the ATLAS$^{\textrm{3D}}$ galaxies in Fig.~\ref{fig:innermassmet}.
The stripping scenario suggests that the resulting objects are outliers in the mass metallicity relation  \citep[see also][]{Chilingarian09}.
The progenitor galaxy follows the mass metallicity relation. The stripping event reduces the stellar mass.
However, the stellar metallicity is not reduced, but remains unchanged or can even be enhanced.
Two things can play a role: first, preferentially stars in the outer parts with lower metallicity will be stripped, leading to higher
averaged values. Secondly,  the interaction  cannot only lead to the stripping 
of stars, but also to gas inflows to the centre where a starburst enabled additional enrichment 
(e.g.~for gas rich progenitors -- as for example was suggested for M32 \citealt{Graham02} -- see also \citealt{Forbes03}).
This possibly can also alleviate an apparent lack of progenitors for the very metal richest CSSs

\subsection{Stripping at work}
Several discoveries of CSSs  caught in the act of formation via stripping
\citep[e.g.][]{Chilingarian09,Huxor11b,Chilingarian15,Jennings15}
undoubtedly tell us that stripping contributes to the population of CSSs.
As seen in  Section \ref{section:stripped_objects}  these `smoking gun' examples share
the characteristic of exceedingly high metallicity when compared to the mass metallicity relation.
This supports the idea that the stripping scenario is a viable option for a large number of CSSs
which have also  metallicities in excess of the expectation from the mass metallicity relation.
The objects  in  Section \ref{section:stripped_objects}  are mostly of relatively young age,
 which seems to be  different from the conventional wisdom of an old age of UCDs and cEs.
However, there is a selection effect in play, since the evidence for a stripping origin
for both the detection of tidal streams and  extended SFHs 
disfavour purely old age.
The detection of tidal streams sets a limit on the age since the interaction, since
these features are rather short lived  \citep[e.g.][]{Rudick09}. When some boost of star formation accompanies the stripping event \citep[e.g.][]{Forbes03,Emsellem08}
 these objects with tidal features are expected to have young average ages. 
Likewise extended SFHs mean that there are  younger stars present, which contribute 
over-proportionally to the light, so that the luminosity weighted SSP-equivalent age cannot be very old.

M60-UCD1 is an exception to this. In this case, the (exceedingly high) mass of the
central black hole suggests a stripping origin \citep{Seth14}. 
Unlike tidal streams, the black hole can be detected 
and its mass measured  also long after the stripping event.
Also, there are several cEs known to host central massive black holes (e.g. NGC~4486B, M32; \citealt{Kormendy97,vanderMarel97}; NGC~5846A has central kinematics very similar to  NGC~4486B based on the spectra with high spatial resolution of \citealt{Davidge08}), which are too massive when comparing to the 
black hole mass spheroid mass relation \citep[e.g.][for M32 it can be discussed whether it follows the relation, or whether its very well measured black hole mass unduly affects and, therefore, biases the relation]{Magorrian98}.
Overly massive black holes can be interpreted as evidences that the CSSs were more massive at an earlier time (just like the exceedingly high metallicities), and  that a stripping event reduced their stellar mass to the observed amount.

\subsection{Isolated cEs}
Interestingly, one of the two isolated CSSs \citep[CGCG 036-042;][]{Paudel14}  has a low metallicity compared to the bulk of our CSSs, which
places it within the scatter of the galaxy mass metallicity relation  (albeit as  an extreme case).
The  other isolated cE in our sample \citep[cE0;][]{Huxor13}, has around solar metallicity.
While the results for metallicity and  [$\alpha$/Fe] are consistent within the uncertainties,
there is some tension for the age, with the age derived from the MODS spectrum being younger than
the \citet{Huxor13} value based on the SDSS spectrum, which has a slightly higher  $S/N$ of 18--27.
We reanalysed the SDSS spectrum and find an age consistent with that of \citet{Huxor13}.
The value  we use throughout the paper is the weighted average, consistently with other
objects that have multiple spectra. This value is consistent within the uncertainties with \citet{Huxor13}.

\citet{Chilingarian15} did not find any statistically significant difference 
between isolated and other cEs.
  \citeauthor{Chilingarian15} concluded that the isolated cE formed in high density regions 
  and then  escaped, avoiding the need for an alternative formation scenario for the rare isolated
  cases  \citep{Huxor13,Paudel14}.

\subsection{GCs, UCDs, and the mass range $\sim$10$^6$--10$^8$M$_{\sun}$}

In the mass range $\sim$10$^6$--10$^8$M$_{\sun}$ CSSs formed by stripping are joined with the high mass end of GCs,
and it has been suggested that UCDs are simply large GCs \citep[e.g.][]{Fellhauer02,Mieske13}.
Recent literature provides evidence that both formation channels, i.e. stripping and large GCs, operate at the low-mass end of the CSS population
(\citetalias{AIMSSI}, \citealt{Hilker06,Brodie11,Chiboucas11,Chilingarian11,daRocha11,Norris&Kannappan11,Pfeffer14}).
Consistent with this mixed  scenario, we observe a wide range of metallicities and old, as well as intermediate,
 ages of our CSSs at low mass.
 
The stellar population characteristics actually can help to tell the stripped objects and star clusters apart.
\citet{Norris15} analysed two of the objects in detail, benefiting from very high quality spectra, and
they were able to constrain the SFHs of the objects. While NGC~3923-UCD1 basically
has only old stars and fits readily into the star cluster category, NGC~4546-UCD1 has a SFH
 extending to the recent past when the progenitor was stripped \citep[see also][]{Norris&Kannappan11}.
Several  other CSSs with similar mass have solar  [$\alpha$/Fe],
including the stripped objects NGC~4546-UCD1 and NGC~3628-UCD1. This could
hint at longer lasting star formation episodes  \citep[e.g.][]{Norris&Kannappan11,Norris15}, 
and it is different from the generally $\alpha$-enhanced Milky Way
GCs \citep{Pritzl05}.  \citet{Evstigneeva08} noted that the reverse is not necessarily valid,
since early stripping can lead to $\alpha$-enhanced UCDs.

 In our sample it appears as if there is a gap in mass between UCDs 
and GCs. However, this is a selection effect,
due to combining (extragalactic) UCDs bright enough to obtain spectroscopy
with Milky Way GCs.
Generally, there is an overlap in mass of UCDs and the GC systems of galaxies, and the possibility of UCDs following the GC
luminosity function was one reason for relating the two \citep{Hilker06,Norris&Kannappan11,Mieske12}.

When comparing to the metallicity distribution  of the extragalactic GCs from \citet{Usher12},
it needs to be considered that their sample also contains
bright objects, which are more massive than $\omega$Cen, since
they did not impose an upper limit in luminosity. 
Of the objects in \citet{Usher12}
10.8\% have super solar metallicities and 6.0\% have $[Z/{\rm H}]>0.2$. When only counting those that 
have 1$\sigma$ larger metallicities than the limits, the fraction reduces to 2.8\% and 1.0\%, respectively.
Some of those massive objects  may be stripped nuclei like 
NGC~3628-UCD1, and should be classified as UCDs. 
Furthermore, it can be expected that none of our CSSs are a GC-type of 
object with very low metallicity ($[Z/{\rm H}]<-1.5$). 
Being restricted to high stellar masses in terms of
GCs means that any GC-like object in the sample is expected
to be affected by the blue tilt, i.e.~should have managed to increase its metallicity 
 due to self-enrichment \citep[e.g.][]{Norris&Kannappan11}.

We note that young massive star clusters (YMCs) in the local Universe reach densities similar to those of the CSSs
 and typically have around solar or even super solar metallicities \citep[e.g.][]{Schweizer&Seitzer98,Maraston01,Maraston04,Strader03,Bastian13}.
 The high metallicities are not surprising in this case, since YMCs are forming from gas in merging spiral galaxies
 at redshift $z=0$. However, this also dictates  their very young age.
 Evolved YMCs may contribute to the population of low-mass CSSs, but seem to be an unlikely option for the old CSSs with highest metallicity.

The more massive CSSs do not span a metallicity range as large as at low mass, and they are all more metal rich than $[Z/\textrm{H}]>-0.5$. 
\citet{Norris&Kannappan11} used the GC luminosity function to estimate the
highest  GC mass expected in the most populous GC systems.
Based on this, no object more massive than a few times $10^7 $M$_{\sun}$ can be a GC.
Interestingly, the CSSs in our sample are exclusively metal rich above a similar mass scale (cf.~Fig.~\ref{fig:massmet}).

The dwarf nuclei of \citet{Paudel10a}, as  objects potentially being liberated by stripping to form CSSs,  
seem to mostly be restricted to sub-solar to at most solar
metallicities and  [$\alpha$/Fe]  in Fig.~\ref{fig:massrad_stellpop}. When considering their full sample, however, this
appears to be a bias effect (see also Fig.~\ref{fig:massmet}) introduced by restricting our sample to 
those objects with measured sizes \citep[from][]{Cote06}.
In their complete sample there are a number of nuclei with slightly super solar metallicities and/or enhanced
 [$\alpha$/Fe].
 \citet{Paudel10a} stated that the dwarf nuclei of those dEs, which are
 located in regions with the same high local projected galaxy density as UCDs,
share similar characteristics with those. This is possibly related to a trend of stripping
being less effective in clusters at later times \citep{Pfeffer14}, and threshed nuclei, as well
as nucleated dwarfs with their star formation ceased long ago, having the tendency to be
located towards the centre of the cluster.

\citet{Francis12} studied a sample of Virgo (including some UCDs also present in  \citeauthor{Paudel10a})
and Fornax cluster UCDs. Their analysis found  also a large spread in metallicities, and mostly old age and super solar
[$\alpha$/Fe]. \citeauthor{Francis12} concluded that the metallicity and age distributions are different from present day nucleated dEs,
and that the one  cannot transform into the other by stripping. Instead, they stated that the UCD metallicity distribution is similar
to that of the GCs, with the UCDs not conforming to a metallicity luminosity relation.
\citet{Brodie11}, on the other hand, described the UCD colour magnitude relation as closely linked to that of  dwarf nuclei, and concluded that the two are likely interrelated.
In Fig.~\ref{fig:massmet}, we show that the  \citet{Paudel10a} sample of nuclei, which were carefully separated from their 
host galaxies, span the complete metallicity range of the UCDs.
Our sample of UCDs contains also some that are younger than classical GCs, as well as objects with solar [$\alpha$/Fe] (Fig.~\ref{fig:massrad_stellpop}),
so that the UCD stellar populations are overall not same as those in GCs.
Moreover, we identified a transition mass, above which the objects are exclusively metal rich, and are not expected to be GCs.
In the context of comparing UCDs with GC and dE nuclei, it is also noteworthy that \citet{Liu15a} found a continuum from
UCDs to UCDs with envelopes to dEs with nuclei, and  \citet{Zhang15} reported  that the spatial distribution of UCDs around M87
and their velocities distribution are distinct from those of the GC system. Both studies concluded that the UCDs are
not exclusively massive GCs.

Additional evidence for a contribution to the CSS population by stripping 
in this mass range was found by
\citet{AIMSSII}, who measured dynamical
masses exceeding the stellar mass significantly \citep[see also][]{Taylor15}. 
They also showed that this enhancement of dynamical mass  was a good fit to the stripping simulations of \citet{Pfeffer13}.
One of these objects with an extreme mass ratio is S999 \citep{Hasegan05,Janz15},
which does not have a very high metallicity ($[Z/\textrm{H}]-0.95$).
 In the picture of UCDs being surviving nuclei of stripped dwarf galaxies
there is a reservoir of nuclei with matching low metallicity. Furthermore, the simulations of \citet{Pfeffer13}
suggest that the remnants can become as small and as low-mass as GCs.

\begin{figure}
\includegraphics[height=0.48\textwidth,angle=-90]{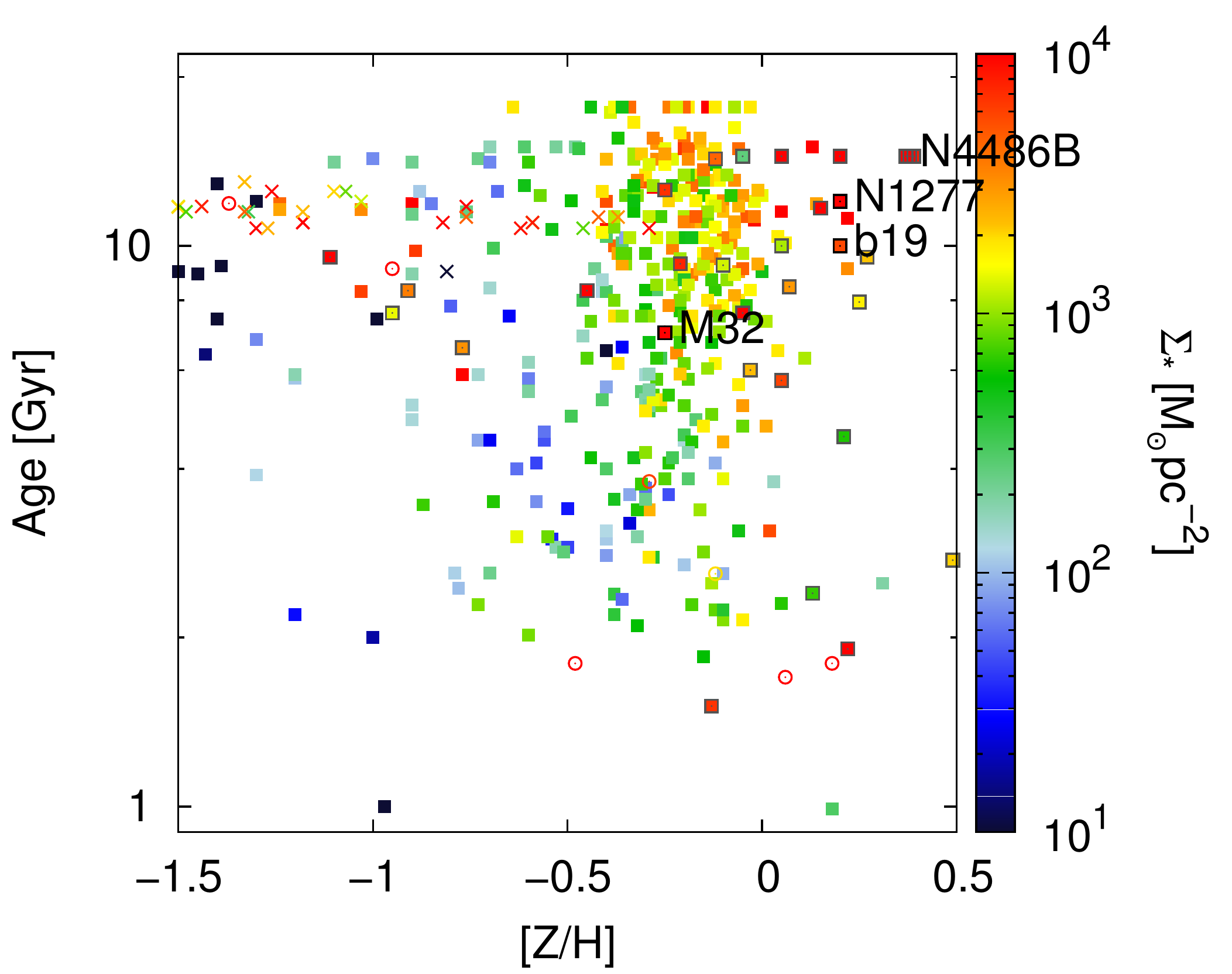}\\
\includegraphics[height=0.48\textwidth,angle=-90]{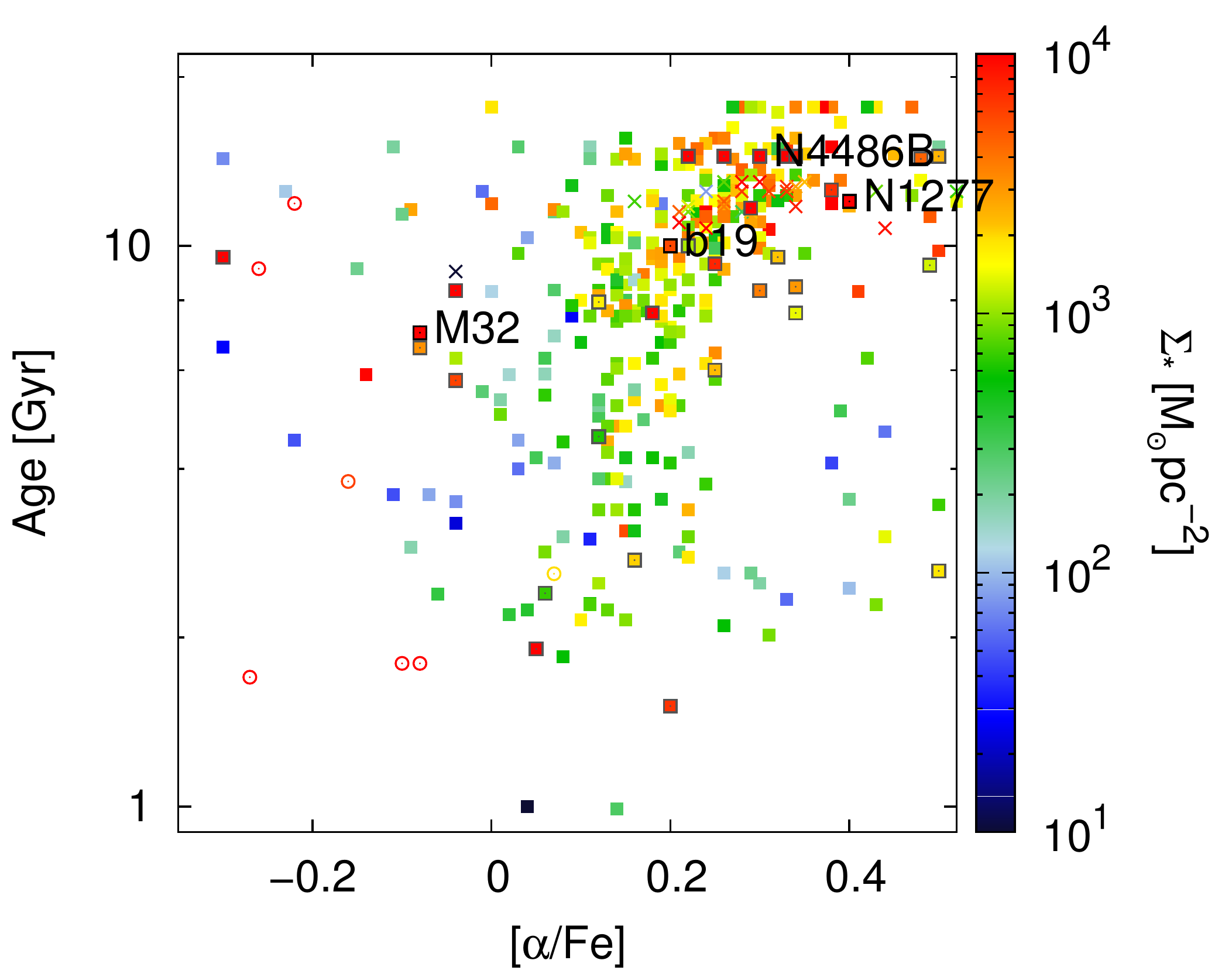}\\
\includegraphics[height=0.48\textwidth,angle=-90]{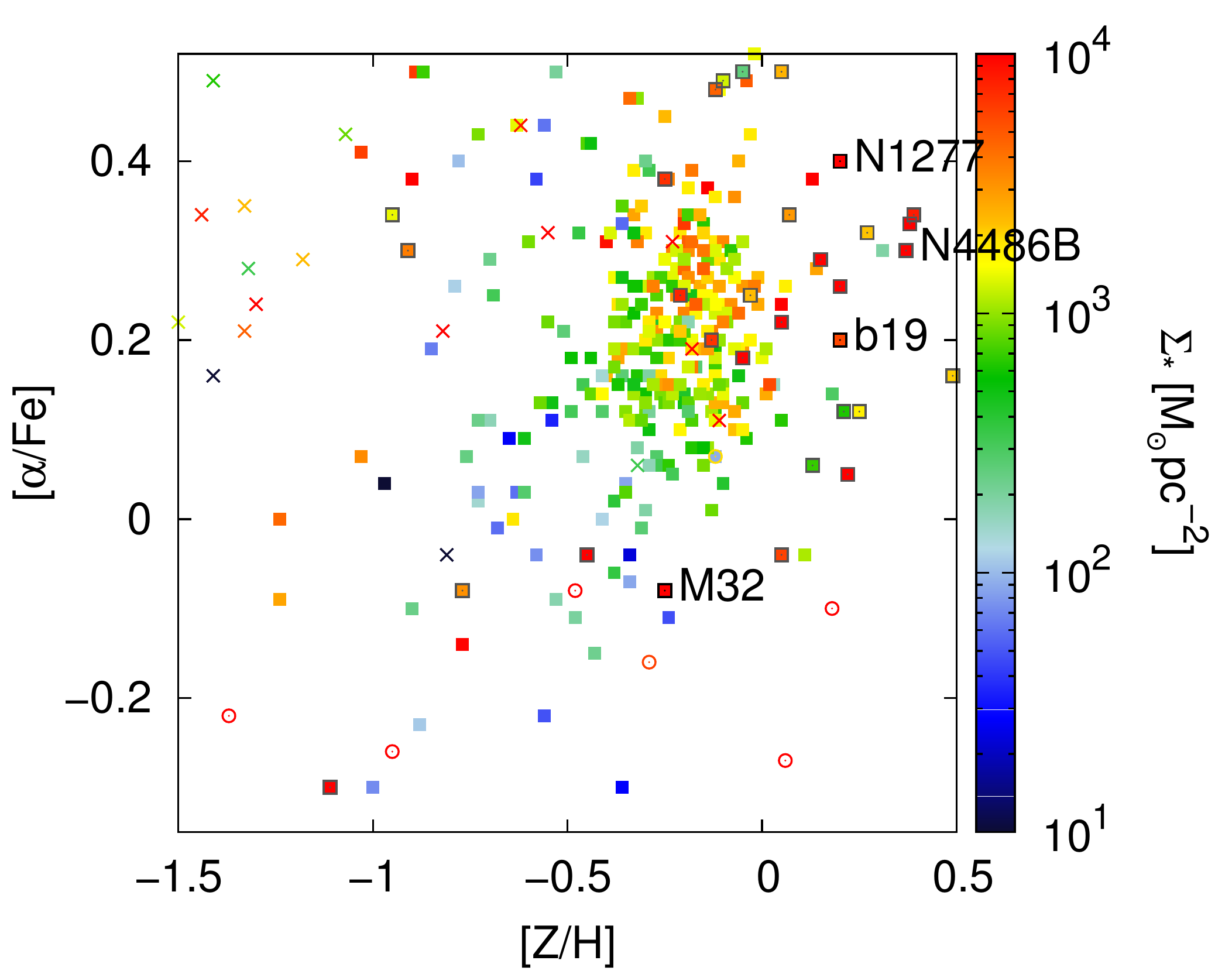}
\caption{From top to bottom: Age versus metallicity, age versus  [$\alpha$/Fe], and  [$\alpha$/Fe] versus metallicity diagrams. The colours display the average surface density within $R_\textrm{e}$, $\Sigma_* = M_*/2\pi  \times  R_{\textrm{e}}^{-2}$. Globular clusters are displayed with crosses, dwarf nuclei with open circles, and the remaining objects with squares. }
\label{fig:stellpop_add}
\end{figure}

\begin{figure*}
\includegraphics[height=0.45\textwidth,angle=-90]{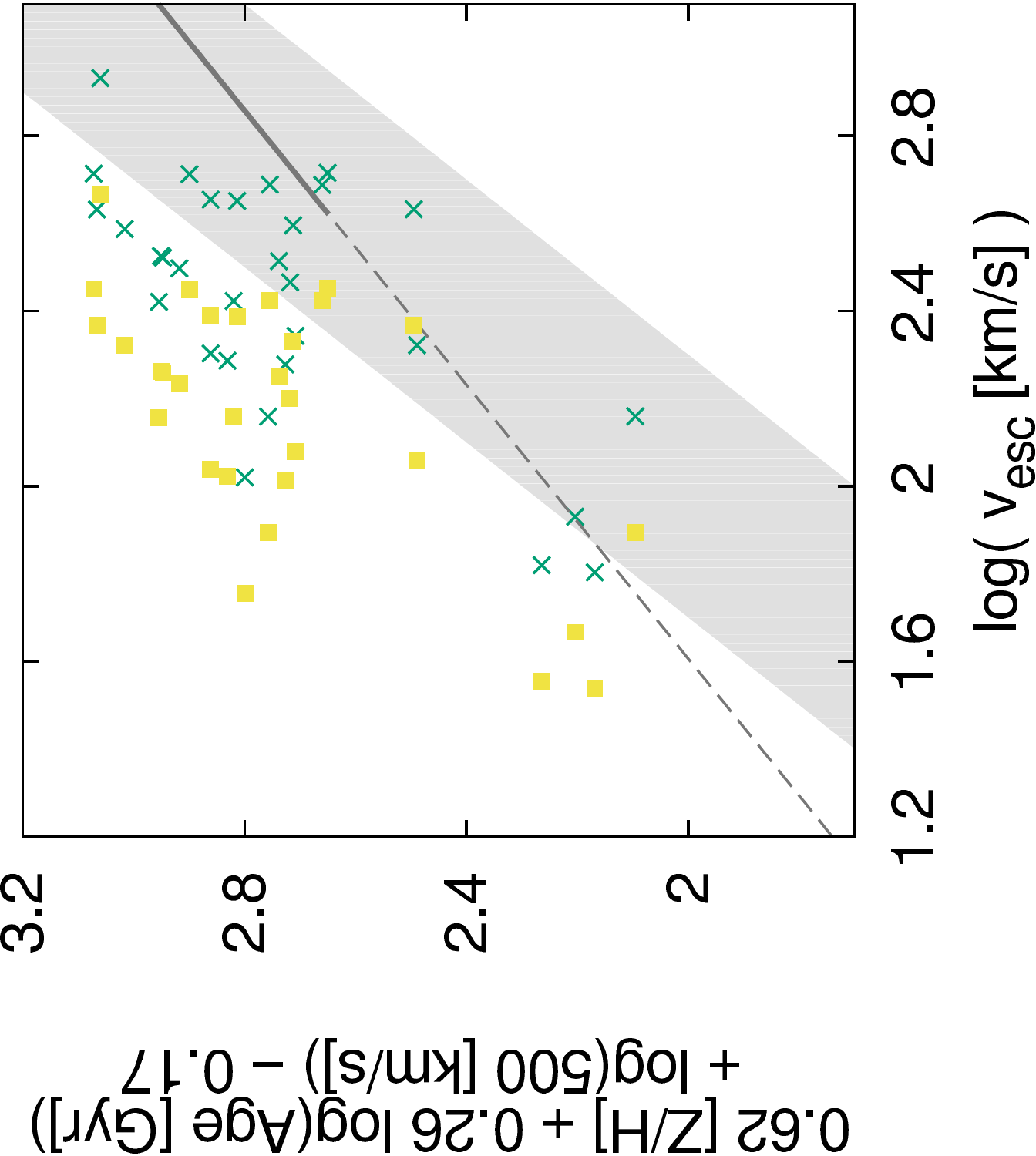}\hskip 1cm
\includegraphics[height=0.45\textwidth,angle=-90]{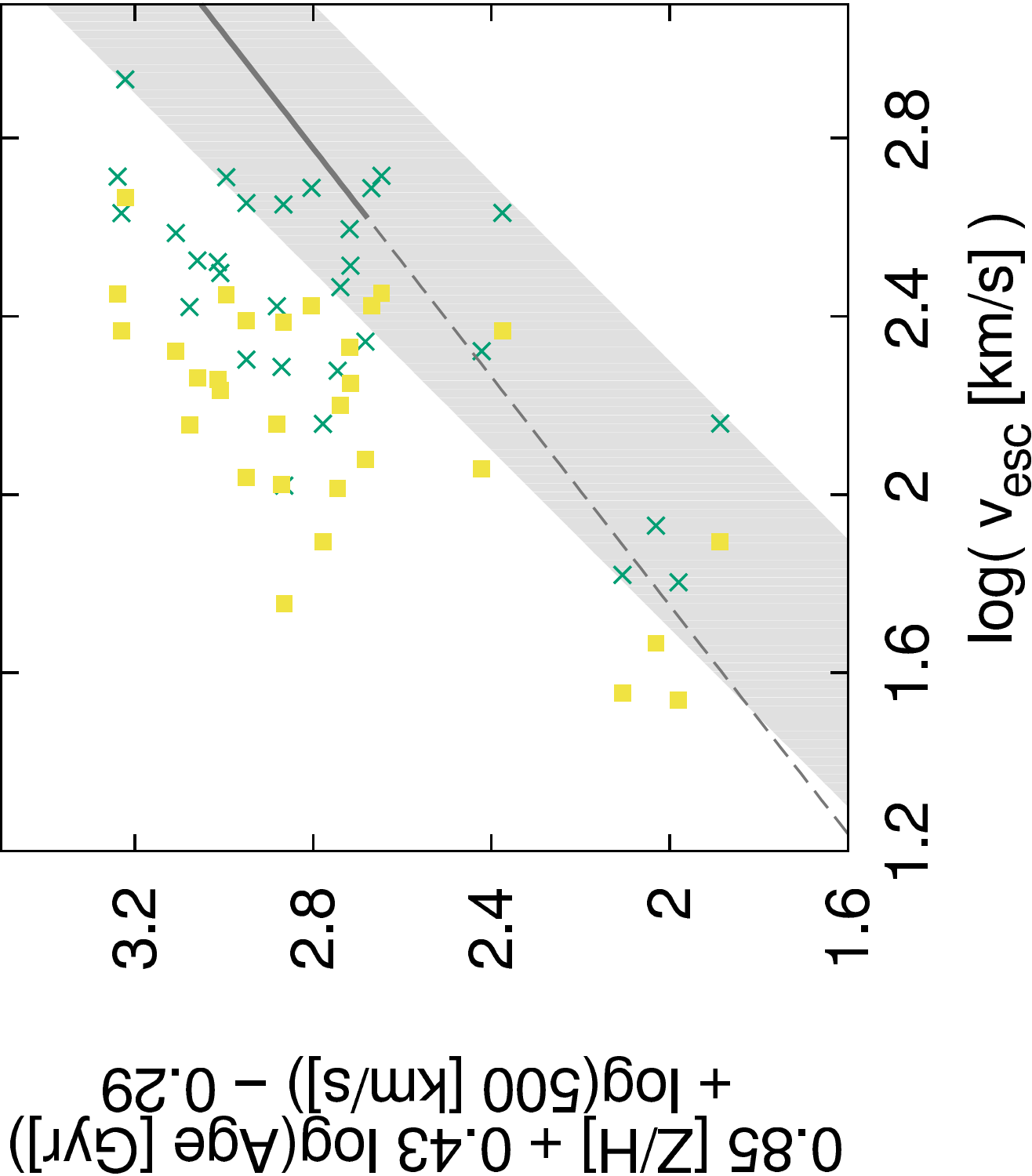}
\caption{SSP-equivalent stellar population parameters and escape velocity.  The two panels show the
edge-on projections of the planes that \citet[left]{ATLAS3DXXI}  and \citet[right]{SAURONXIV}  found to be
spanned by the local stellar population parameters and escape velocities in early-type galaxies.
The grey shaded regions indicate these planes $\pm0.3$ dex, which is chosen by eye to generously  
describe regions that were populated by the profiles in these studies.
For our CSSs we calculate the corresponding linear combinations of luminosity weighted age and metallicities
and lower (yellow boxes) and upper  (green crosses) limits for the luminosity weighted escape velocities (see text).
The grey lines indicate the relations found by \citet{ATLAS3DXXI} and \citet{SAURONXIV} for escape velocities
above 400 km s$^{-1}$, where the authors found the scatter to be considerably smaller. The CSSs fall on average
well above the planes and relations for early-type galaxies, suggesting that their enrichments exceed what is 
possible for their current potential wells.\label{fig:v_esc}}
\end{figure*}

\subsection{Compact ellipticals and the mass range of $\sim$10$^8$--10$^{\textbf{10}}$ M$_{\sun}$}

At  higher masses of $\sim$10$^8$--10$^{\textrm{10}}$ M$_{\sun}$, the high metallicities  (in  comparison  to the general mass--metallicity relation)
 potentially indicate that the objects were more massive galaxies in their past, and only later reduced in mass
by stripping of material \citep[see also e.g.][]{Chilingarian09}.
The two of our objects that have been observed to feature tidal tails \citep{Huxor11b}
strengthen this scenario. One of the  objects is comparably young, which may be
expected if stars were formed in, or up until, the interaction, which happened rather recently.
There are 9 additional cEs known to be embedded in tidal streams  \citep{Chilingarian09,Chilingarian15}.
These objects show a range of age (6.3 Gyr and older), with four of them being older than 10 Gyr.
Their metallicities are generally high ($-0.34\le[\textrm{Fe}/\textrm{H}] \le 0.12$), with most of them being around solar
and higher. In particular, the latest study \citep{Chilingarian15} increases the number of cEs in tidal streams substantially, 
 thereby amplifying  the mounting evidence for cE formation via tidal stripping.
Additional evidence for a past as a more massive galaxy includes 
overly-massive central black holes, such as  NGC~4486B, in which the black hole contributes 11\% 
to the total mass -- much more than the 0.1\% expected from the black hole bulge mass 
scaling relation \citep{Magorrian98}. 
Sometimes, also a two-component structure, such as that seen in M32 \citep{Graham02}, was taken 
as evidence of a stripped disc.

Several of the high mass CSSs in the literature have more normal metallicities, i.e.~sub-solar.
For some of them  [$\alpha$/Fe] was not measured. Thus the comparison
might not be entirely fair, since 
including  [$\alpha$/Fe] in the metal budget can bias the metallicity result to higher values.
The other  three low metallicity cEs are in the Virgo cluster and they represent the least compact
objects in the sample of \citet{Guerou15}.
Their size is large enough that they may potentially be regarded as small \emph{normal} dEs.
 
At even higher masses there are some compact galaxies that have recently acquired a lot of attention.
They   have very high velocity dispersions for their luminosity, and they are close to or even
within the \emph{zone of avoidance} (Fig.~\ref{fig:massrad}).
 \citet{vandenBosch12} claimed that NGC~1277 also has an overly massive central black hole,
 with the most extreme value for the mass contribution of the black hole at  59\% of the bulge mass.
While \citet{Lasker13} do not have definite proof for an overly massive black hole in 
J151741.75-004217.6 or b19, they consider it likely.

These objects also share very high metallicities with our CSSs.
There are also potentially more of these kinds of objects \citep{Saulder15}.
While the characteristics of high metallicity and overly massive black holes could
inspire similar ideas for their formation, their origin is thought to be different.
In part this is due to a lack of potential hosts to cause the stripping.
Instead, it has been suggested that they constitute descendants of
compact galaxies at high redshift, so-called \emph{red nuggets} \citep{vanderWel14}.
In this context, it is very interesting that recently \citet{Lonoce15} measured a very high metallicity ($[Z/\textrm{H}]>0.5$) 
for an early-type galaxy at redshift $z$$\sim$1.4.
While unevolved red nuggets may be expected to have very old ages, unlike some
of our CSSs, it is beyond the scope of this paper to explore whether the formation channels 
for red nuggets are a possibility for some other of CSSs \citep[c.f. also][]{Graham15}. 
However, the possibility reminds us that stripping is not
necessarily the only formation mechanism for CSSs, and some of them may in a sense
be  the \emph{true continuation} of ellipticals towards lower mass \citep[e.g.][]{Kormendy09}.

\subsection{Enrichment and escape velocity}

It is unclear whether the mass metallicity relation is caused by a primary dependence of metallicity on mass, 
or whether it is a consequence of fundamental relationships between other quantities, 
such as velocity dispersion or escape velocity \citep[e.g.][]{Davies93,Bernardi03}, which are also related to mass.
In Fig.~\ref{fig:massrad_stellpop} lines with constant velocity dispersions, assuming the virial theorem
and constant virial coefficient, run somewhat steeper than the border of the zone of avoidance. 
The metallicities of all objects overall
seem to trace  velocity dispersion much more closely than stellar mass \citep[c.f.][]{Guerou15}.
In Fig.~\ref{fig:stellpop_add}, the parameter space spanned by age, metallicity, and  [$\alpha$/Fe]
is shown with the colours of the symbols displaying the effective stellar surface density within $R_{\textrm{e}}$.
Systematic trends in these plots suggest higher dimensional relations of the parameters involved 
\citep[e.g.][]{Brodie11,Guerou15,Sandoval15}. Instead of exploring those in depth, we focus in the following
on relations with escape velocity.

\citet{SAURONXIV,ATLAS3DXXI} made use of dynamical modelling and metallicity maps
from the SAURON IFU to show that metallicity and escape velocity are 
related  locally within the galaxies  for a substantial sample of early-type galaxies. Especially when using Mg$b$ 
as a proxy for metallicity, they found a very tight relation. Moreover, the authors discovered that their early-type galaxies span a plane 
in the four-dimensional space of escape velocity and SSP-equivalent stellar population parameters age, metallicity, and
 [$\alpha$/Fe]. Again, this is also true for the local escape velocity and stellar population characteristics.

For our CSSs, spatially resolved spectroscopy, needed for dynamical modelling and determination of stellar population parameters locally, is unavailable.
Instead, we compare their global parameters to the findings of \citet{SAURONXIV,ATLAS3DXXI} in Fig.~\ref{fig:v_esc}.
For the escape velocity we use two different methods, which are designed to be lower and upper limits. 
The stellar population parameters from the spectra are light-weighted averages. 
In order to find the corresponding light-weighted average escape velocity, we first assume a constant mass-to-light ratio and
a Plummer sphere \citep{Plummer11} for convenient integration. With the stellar mass and size of
the CSSs the value within the projected half-light radius can be readily calculated. However, in particular for the cEs, 
this will underestimate the real light weighted escape velocity, since their profile is steeper
and both light and escape velocity increase towards the centre. Therefore, we calculate as a second method
an upper limit by assuming a Dehnen profile \citep{Dehnen93} with $\gamma=3/2$, which approximates the
de-projected mass profile for a de Vaucouleurs profile, and find its maximum escape velocity. Note that this should
represent a generous upper limit, since we use the maximum escape velocity, and since the majority of CSSs 
including the cEs have profiles shallower than a de Vaucouleurs profile (which corresponds to a S\'ersic profile with an index $n=4$, while most
CSSs have $1<n<2.5$).

Fig.~\ref{fig:v_esc} illustrates that even for this latter maximum escape velocity the CSSs, and especially the cEs, fall partly beyond the regions traced by the early-type galaxies. This is true for both the area  enclosing all the local values of  \citet{SAURONXIV,ATLAS3DXXI} and
the relations the authors found for escape velocity $v_\textrm{esc} > 400$ km s$^{-1}$, which is even tighter according to them.
At lower escape velocity the scatter is increased, mostly by galaxies with negative gradients and central star formation and dust.
We include the comparison to both studies, since \citet{ATLAS3DXXI} pointed out that their sample, although larger, 
included poorer quality data, so the earlier work might provide a more reliable reference plane. The cEs fall above the relations
in both cases.
If the local escape velocity indeed determines the effectiveness of the enrichment process, the exceedingly high metallicities 
of the CSSs (implying $v_\textrm{esc}\gtrsim600$ km s$^{-1}$ using the planes) are a strong indication that those objects were more massive in their past.

\section{Conclusions}
\label{section:summary}

We have presented the stellar population characteristics (single stellar population equivalent age, metallicity, and  [$\alpha$/Fe]) for a sample of 29 compact stellar systems (CSSs)
and compared to literature values for CSSs and other types. Many of the CSSs have metallicities that are \emph{too high} when compared to the mass metallicity relation traced by the comparison sample of early-type galaxies. In fact some of the objects appear to have a metallicity exceeding that of the inner parts (within $R_\textrm{e}/8$) of massive early-type galaxies.

At high mass, the departure  of the cEs  from the galaxy mass--metallicity relation argues against them being a simple continuation of the massive early-type galaxies.
In the UCD mass range, we find a transition such that the metallicity distribution of objects less  massive than a few times $10^7$ M$_{{\sun}}$ is wide, similar to that of GCs,
while more massive objects are all metal rich. This transition coincides with the mass at which luminosity function arguments previously suggested the GC population ends.

The  high  metallicities in UCDs and cEs are only paralleled by those of dwarf nuclei and the central parts of massive early types.
We interpret these findings  as an indication that they were more massive at an earlier time and underwent tidal stripping to obtain their current mass and compact size. This  interpretation is supported by the fact that objects with direct evidence for a stripping origin have similarly high metallicities. 
 
 Future instrumentation will provide the means to hunt for further clues about the formation histories of CSSs,
 such as detecting and weighing supermassive central black holes. Other promising avenues include the
 taxonomy of individual elements in the stellar populations \citep[e.g.][]{Evstigneeva07,Taylor10,Colucci13,Strader13}
 and exploring multi-dimensional relations between the stellar populations and physical parameters such as mass, size, and
 compactness as hinted at by Fig.~\ref{fig:stellpop_add} \citep[e.g.][]{Brodie11,Guerou15,Sandoval15}.

For now, the integrated stellar population parameters offer valuable clues to the formation history of CSSs, especially since such information is considerably easier to obtain than resolved star formation histories or direct measurements of overly massive central black holes.
  Young age and solar  [$\alpha$/Fe],  which are measurable in objects where  tidal tails have already faded, can provide evidence to distinguish them  from massive star clusters.

Furthermore, we followed the studies of \citet{SAURONXIV,ATLAS3DXXI} in comparing the stellar populations as a function of escape velocity. The authors found 
 massive early-type galaxies to form a plane in a  parameter space spanned by  the local escape velocity and the local stellar population parameters, i.e.~age, metallicity, and   [$\alpha$/Fe]. A plausible explanation could be that the efficiency of enrichment depends locally on the ability to retain metals, as indicated by the escape velocity. The CSSs  fall, on average, above these narrow planes found by \citet{SAURONXIV,ATLAS3DXXI}, which can be understood  as evidence that their current mass is too small for the level of enrichment that the CSSs have reached. This is a strong argument in favour of the stripping scenario, and suggests that  metallicity can be utilized to tell apart objects with a more massive past in the transition region of massive objects with sizes of around 500pc.

\section*{Acknowledgements}
We thank our referee Igor Chilingarian for useful suggestions that helped improving the paper.
We thank Carlos Morante for helpful discussions. JJ and DAF thank the
ARC for financial support via DP130100388.
This work was supported by
NSF grants AST-1109878,  AST-1515084, and AST-1518294.
MJF gratefully acknowledges support from the DFG via Emmy Noether Grant Ko 4161/1.

This paper used data obtained with the MODS spectrographs built with
funding from NSF grant AST-9987045 and the NSF Telescope System
Instrumentation Program (TSIP), with additional funds from the Ohio
Board of Regents and the Ohio State University Office of Research.

The LBT is an international collaboration among institutions in the United States, Italy and Germany. LBT Corporation partners are: 
The University of Arizona on behalf of the Arizona university system; Istituto Nazionale di Astrofisica, Italy; LBT Beteiligungsgesellschaft, 
Germany, representing the Max-Planck Society, the Astrophysical Institute Potsdam, and Heidelberg University; The Ohio State University, 
and The Research Corporation, on behalf of The University of Notre Dame, University of Minnesota and University of Virginia.

Based on observations obtained at the Gemini Observatory, as part of programs GS-2011A-Q-13, 
GS-2013A-Q-26, GS-2014A-Q-30, and processed using the Gemini IRAF package. The Gemini 
Observatory is operated by the Association of Universities for Research in Astronomy, 
Inc., under a cooperative agreement with the NSF on behalf of the Gemini partnership: the 
National Science Foundation (United States), the National Research Council (Canada), 
CONICYT (Chile), the Australian Research Council (Australia), Minist\'{e}rio da Ci\^{e}ncia, 
Tecnologia e Inova\c{c}\~{a}o (Brazil) and Ministerio de Ciencia, Tecnolog\'{i}a e Innovaci\'{o}n 
Productiva (Argentina).

Some of the data presented herein were obtained at the W.M.~Keck Observatory, which is 
operated as a scientific partnership among the California Institute of Technology, the University 
of California and the National Aeronautics and Space Administration. The Observatory was 
made possible by the generous financial support of the W.M.~Keck Foundation. 

The authors wish to recognize and acknowledge the very significant cultural role and reverence 
that the summit of Mauna Kea has always had within the indigenous Hawaiian community.  
We are most fortunate to have the opportunity to conduct observations from this mountain. 

Funding for SDSS-III has been provided by the Alfred P.~Sloan Foundation, the Participating Institutions, the National Science Foundation, and the U.S.~Department of Energy Office of Science. The SDSS-III web site is http://www.sdss3.org/.

SDSS-III is managed by the Astrophysical Research Consortium for the Participating Institutions of the SDSS-III Collaboration including the University of Arizona, the Brazilian Participation Group, Brookhaven National Laboratory, Carnegie Mellon University, University of Florida, the French Participation Group, the German Participation Group, Harvard University, the Instituto de Astrofisica de Canarias, the Michigan State/Notre Dame/JINA Participation Group, Johns Hopkins University, Lawrence Berkeley National Laboratory, Max Planck Institute for Astrophysics, Max Planck Institute for Extraterrestrial Physics, New Mexico State University, New York University, Ohio State University, Pennsylvania State University, University of Portsmouth, Princeton University, the Spanish Participation Group, University of Tokyo, University of Utah, Vanderbilt University, University of Virginia, University of Washington, and Yale University.

\appendix

\section{Stellar population parameters and stellar masses}

\begin{figure*}
\centering
\includegraphics[height=\textwidth,angle=-90]{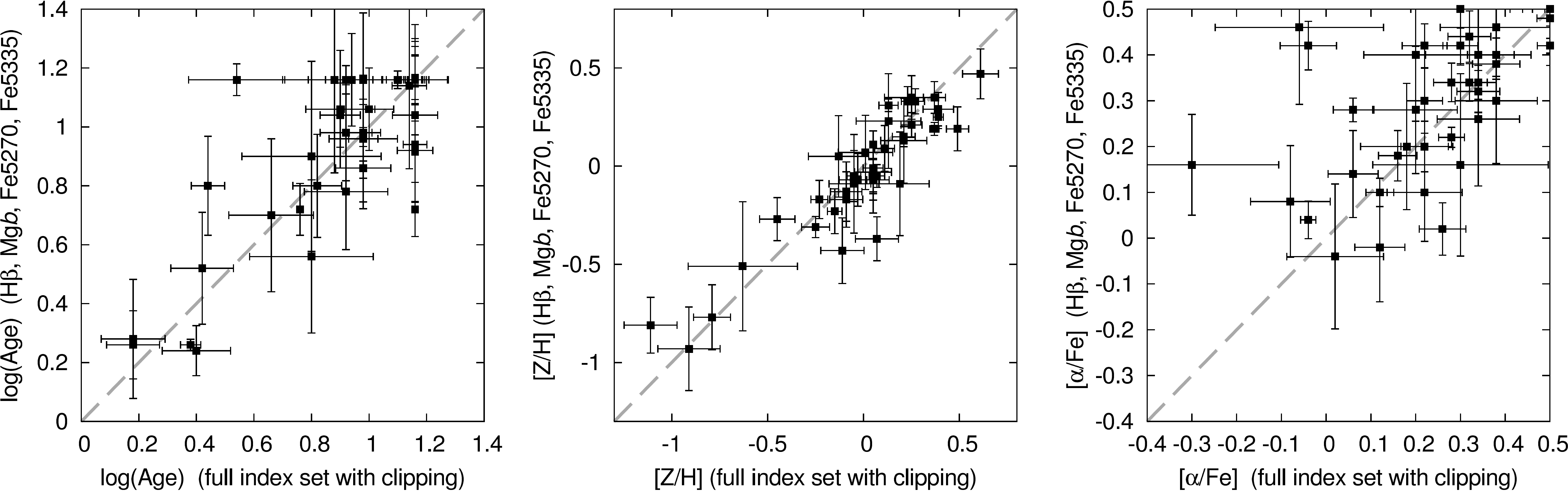}
\caption{Comparison of the stellar population parameters obtained by fitting the SSP models to different index sets as described in Section \ref{section:stellpop}. The grey lines show the 1:1 relations. }
\label{fig:stellpop_comp}
\end{figure*}

\begin{figure*}
\centering
\includegraphics[height=\textwidth,angle=-90]{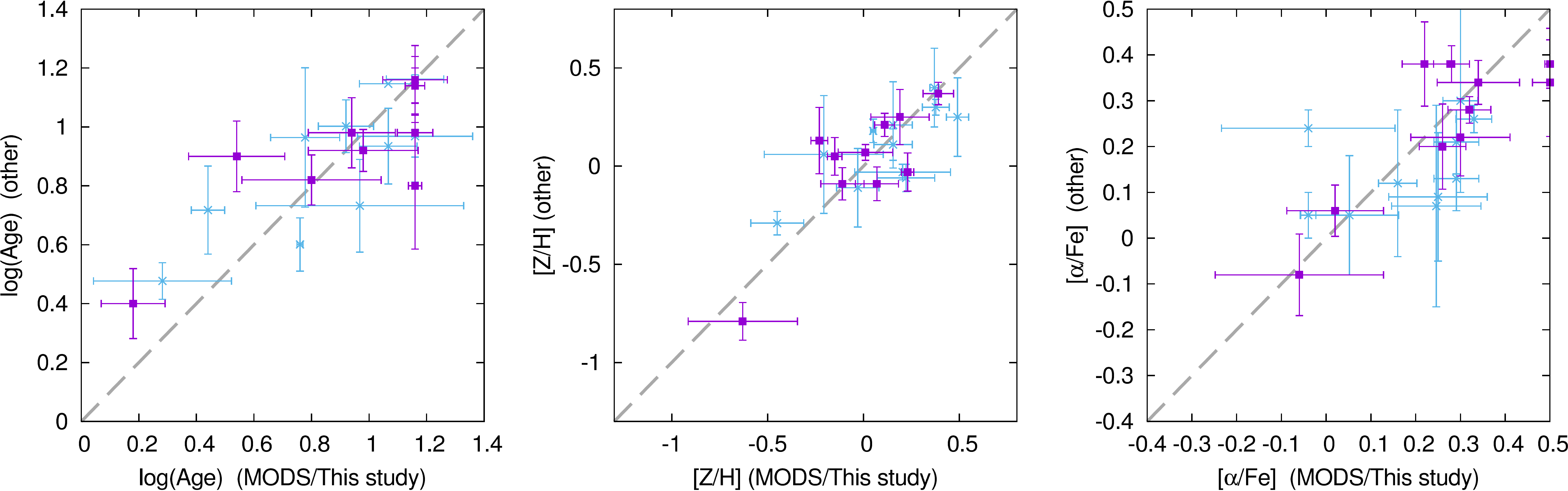}
\caption{Comparison of the stellar population parameters obtained with different instruments within this study (violet boxes) and to literature (blue crosses).  The grey lines show the 1:1 relations.} 
\label{fig:comp_intlit}
\end{figure*}

\begin{figure}
\centering
\includegraphics[height=0.4\textwidth,angle=-90]{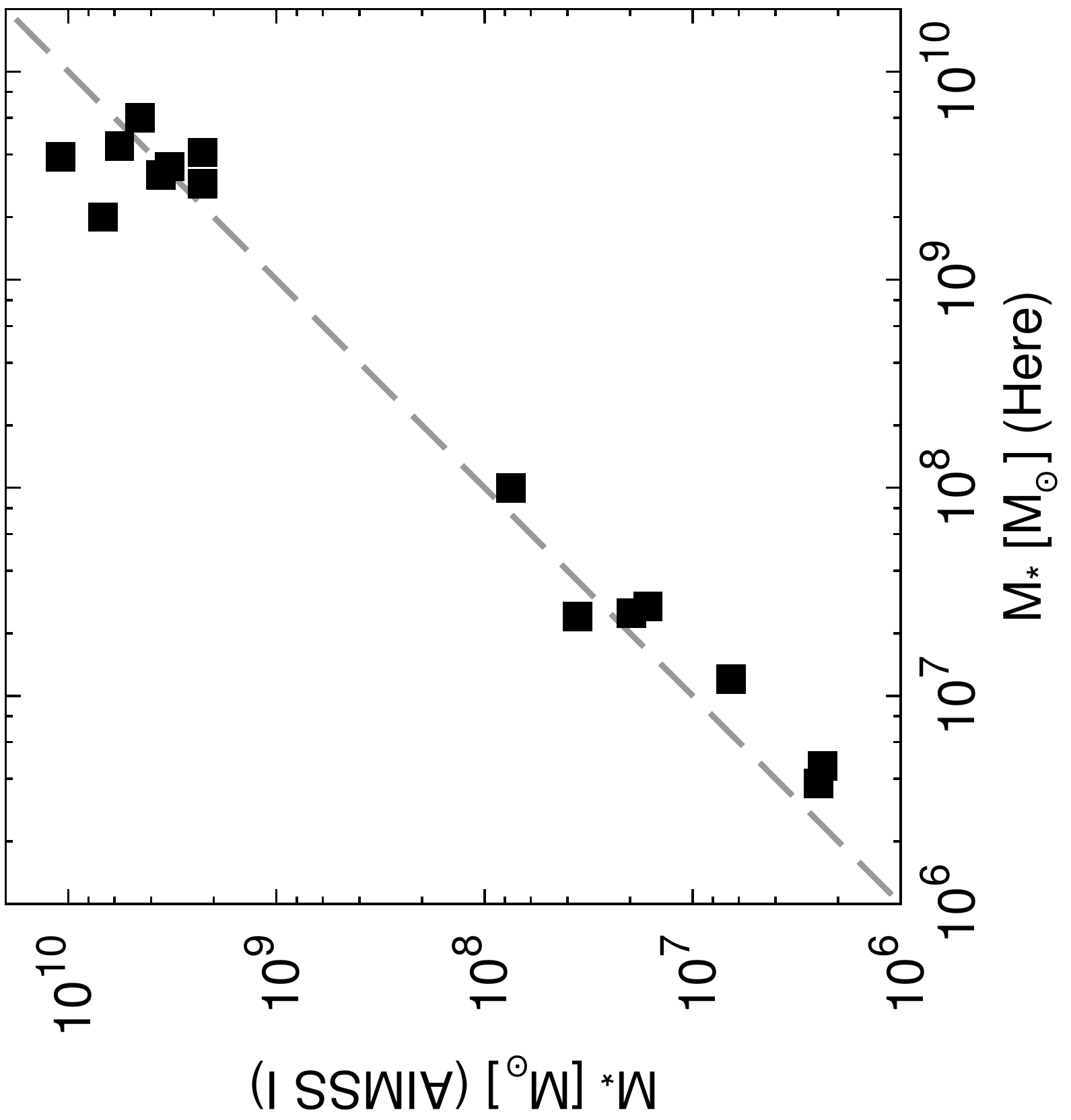}
\caption{Comparison of the stellar masses obtained from the spectroscopic stellar population parameters here and those in the \citetalias{AIMSSI} catalogue, which were
 determined with broad-band photometry in multiple broad-band filters.  The grey line shows the 1:1 relation.} 
\label{AIMSSmass}\label{lastpage}
\end{figure}

We derive the stellar population parameters in two different ways. The full set of indices was used by iteratively clipping indices that did not fit the best-fitting model
and finding the best model with the remaining ones. Alternatively, we used a small index set (H$\,\beta$, Mg$b$, Fe5270, and Fe5335) similar to ATLAS$^\textrm{3D}$ \citep{ATLAS3DXXX}. In Fig.~\ref{fig:stellpop_comp} the comparison of the two is shown.
Furthermore, we provide in Fig.~\ref{fig:comp_intlit} a comparison for those objects with two different spectra. For four objects we analysed the SDSS spectra in addition with the same procedures (M85-HCC1, M59cO, cE0, cE2). Also, for several objects (cE0, cE1, cE2, M59-UCD3, M59cO, M60-UCD1, M85-HCC1, NGC~3923-UCD1, NGC~4546-UCD1, NGC~4486B) a comparison of our adopted values to the literature is shown \citep{Sanchez-Blazquez06,Chilingarian08b,Huxor11b,Huxor13,Strader13,Liu15b,Norris15,Sandoval15}. 
Given the variety of literature sources with differences in data quality as well as in the stellar population models used for the fitting, the agreement is good. In particular, the metallicities are reliably constrained. 

In Fig.~\ref{AIMSSmass} our stellar masses are compared to those in the \citetalias{AIMSSI} catalogue.
Our  stellar masses are derived using  the spectroscopic SSP stellar population parameters obtained here.
This way we can homogenously calculate the masses for all  our CSSs and do not need to rely on
the heterogenous sets of available photometry. The stellar masses obtained in this way generally show
good agreement with those in \citetalias{AIMSSI}. Those  are based on multi-band photometry, 
but allowing for a composite stellar population with a young and  an old component.

\bsp	

\end{document}